\documentclass[twocolumn,superscriptaddress]{revtex4-1}
\usepackage{lipsum}
\usepackage{dcolumn}
\usepackage{bm}
\usepackage{color}
\usepackage{physics}
\usepackage{mathtools}  
\usepackage{xfrac}  
\usepackage{graphicx}
\usepackage{listings}
\usepackage{minted}
\usepackage{amsmath, amsthm, amssymb, calrsfs, wasysym, verbatim, bbm, geometry} 
\newcommand{\beq}{\begin{equation}}
\newcommand{\eeq}{\end{equation}}

\def\ltape{\hbox{\ $<$\hskip -8pt\raise -4pt\hbox{$\sim$}\ }}
\def\gtape{\hbox{\ $>$\hskip -8pt\raise -4pt\hbox{$\sim$}\ }}

\begin{document}

\title{Linear Tearing Growth and Onset of Relativistic Magnetic Reconnection in the Presence of Shear Flows and a Guide Field}

\author{Sarah Peery}
\affiliation{Dartmouth College, Hanover, NH 03750}
\author{Yi-Hsin~Liu}
\affiliation{Dartmouth College, Hanover, NH 03750}

\date{\today}

\begin{abstract}
It has been shown in non-relativistic tearing theory that shear flows will slow the linear phase of tearing instability and can delay onset of magnetic reconnection. We find using kinetic particle-in-cell simulations that shear flow as well as guide field strength affect the onset time of relativistic magnetic reconnection. To model this we develop a numerical solver for the growth rate of the relativistic linear tearing instability, including effects of the motional electric field which has not previously been done. We find slowing of growth due to both shear flows and guide field, and at higher flow shear, transition through an intermediate regime to linear Kelvin-Helmholtz instability.
\end{abstract}

\maketitle

\section{Introduction}
Magnetic reconnection is an energetic process ubiquitous to many plasma systems, wherein magnetic field line connectivity is rearranged, releasing magnetic energy into plasma kinetic and thermal energy. In particular, magnetic reconnection in the relativistic regime, where magnetic energy density is larger than the plasma rest energy density, has been the subject of recent studies. These conditions are relevant to high energy astrophysical objects such as black hole accretion disks and pulsar wind nebula, where reconnection may be responsible for non-thermal particle acceleration 
\citep{lyubarski_01, Lyutikov_2002, benoit_12, drenkhahn_02, guo_15, hoshino_12, kirk_03, lyutikov_03}. In this work, we focus on the effects of an out-of-plane guide field and sheared plasma flows parallel to the reconnecting component of the magnetic field, such as can be found in jets with helical magnetic fields \citep{boccardi_16, wang_21, coroniti_90, mbarek_22}. 

The relativistic regime is characterized as where magnetic energy density is larger than the plasma rest energy density, or in terms of the magnetization parameter, $\sigma= B^2/h>1$ where $h$ is the plasma enthalpy, $h= \rho c^2+ [\gamma_c/(\gamma_c-1)] P$. Here $\rho$ is mass density, $P$ is the pressure in the fluid proper frame, and $\gamma_c$ is the ratio of specific heats. 
In relativistic plasmas, the total Alfv\'en velocity is $V_A/c= \sqrt{\sigma/(1+\sigma)}$, which approaches the speed of light for large magnetization. This, as well as the inclusion of high-speed shear flows and out-of-plane drift necessitates the consideration of special relativistic effects \citep{Kagan_15, uzdenski_10, hoshino_12}. 

In the relativistic regime, the shear flows and guide field introduce additional behavior.
The relevant quantity for reconnection is the projection of the total Alfv\'en speed into the outflow direction $V_A B_{rec}/|B|$, or $V_{A, rec}/c= \sqrt{\sigma_{rec}/ (1+ \sigma_{rec}+\sigma_g)}$, where $\sigma_{rec}$ and $\sigma_g$ are the $\sigma$ based on the reconnecting and guide field components, respectively Therefore, unlike in the non-relativistic limit, the presence of a guide field will affect the Alfv\'en velocity \citep{peery} and reconnection rate.
Further, in this regime the motional electric field, especially in the presence of fast flows, can be of the same order of magnitude as the magnetic field and cannot be neglected in the force balance. 

In both the non-relativistic and relativistic regimes it has been well established that shear flows parallel to the reconnecting field stabilize reconnection and, above a critical velocity (related to $V_A$), suppress reconnection altogether. This has been found for a range of initial conditions including Harris sheets, double current sheets and plasmoid instability \cite{mahapatra, hosseinpour_18, figuerdo, voslion, peery}. 
Several simulation studies find that, if the shear width of the flow is greater than the magnetic shear width, the flow can have an accelerative effect instead \cite{li_10, lakhina, wu14}, however we do not consider this parameter regime.

This paper extends the studies of Peery et al. \cite{peery}, which examined the steady state scaling of relativistic reconnection in pair plasmas under the presence of shear flow ($V_{sh})$ and guide field ($B_g$). A model for the outflow velocity of the reconnection jet was derived from the force balance, including the motional electric field and it was found that the shear speed at which reconnection is suppressed is strongly related to $V_{A, rec}$.
\begin{figure}[h]
  \centering
  \includegraphics[width=\linewidth]{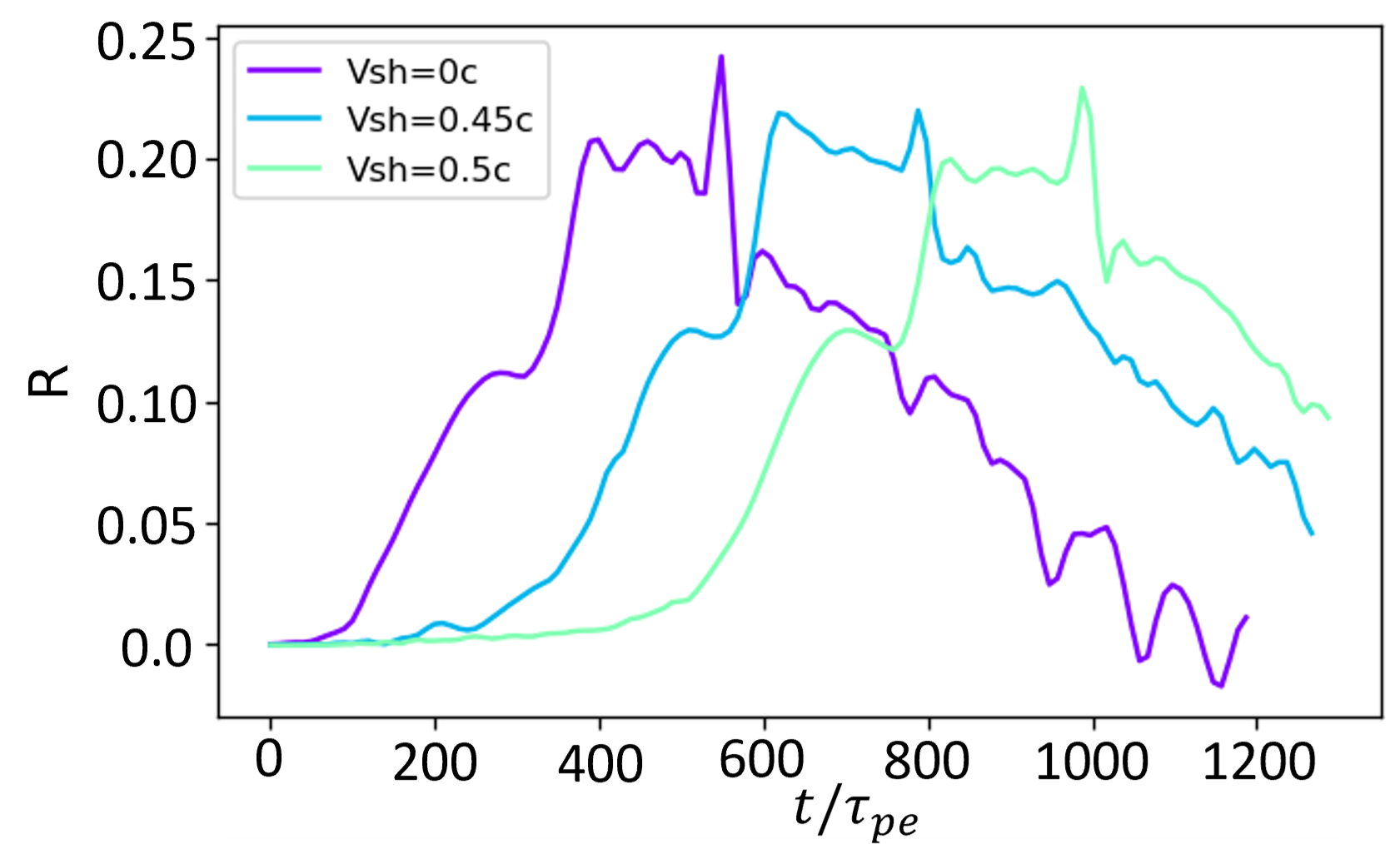}
  \caption{Reconnection rates as a function of time from simulations (used in \cite{peery}) with $B_g=0.01$ and different $V_{sh}$. This show that shear flow speeds delay the onset of reconnection, a behavior we seek to explain in this paper.}
  \label{zero} 
\end{figure}

Several other trends were also noted in \cite{peery}, out of the scope of their analysis, which focused on steady state quantities. Specifically, the simulations demonstrated a delay in the onset time of reconnection and slowing of evolution, depending on shear flow speed. This behavior can be seen in Fig. \ref{zero}, which shows the evolution of the reconnection rates for three simulations with increasing shear flows, results of which were previously used in \cite{peery}. Clearly the time of reconnection onset depends on the shear flow velocity. 

Since the onset and trigger of magnetic reconnection is still an ongoing area of research, this study is motivated to examine factors that affect the time development of the system.
Reconnection starts with a slow buildup phase, as the initial fluctuation grows to a point where it destabilizes the current sheet to form an X-point, starting fast non-linear growth. The trigger may be a number of phenomena, including the KHI or LHDI \cite{nakamura_13, Chen_97, Lu_2013}, but most typically it is the tearing instability \cite{Uzdenski16, dongkuan, Betar, loureiro, honshino21, Tenerani, fkr, coppi}. So in this study we focus on the linear growth of tearing.

The linear tearing instability in the presence of a shear flow has been well studied in the non-relativistic regime.
Seminal work was done in the resistive regime by Chen and Morrison \cite{chenmorrison} who found the growth rate scales with a power of the parameter
\beq
\label{one}
1-C_0\bigg[\frac{\partial_z G(0)}{\partial_z F(0)}\bigg]^2 
\eeq
where $F(z)=\vb{B}_{0}(z)\cdot \hat{k}/|B|$ and $G(z)=\vb{V}_{sh}(z) \cdot \hat{k}/V_A$ and $C_0$ is a constant.
When  $[\partial_z G(0)/\partial_z F(0)] <1$ tearing is totally stabilized.
Subsequent studies of resistive and inertial tearing in both the constant and non-constant $\psi$ regimes support this, finding that the growth rate as a function of shear flow will scale with the parameter in Eq. (\ref{one}) for $C_0$ determined by initial conditions. \cite{jonghe, ofman, mallet, Mallet_2025, faganello}.

In the relativistic fluid description, Yang \cite{Yang_2019} derived the linear growth with both resistivity and inertial effects and found scalings which limit to non-relativistic oblique tearing growth rate \cite{baalrud}. 
However, these relativistic studies have not included electric field terms. We expect the motional electric field induced in the system by the shear flows and guide field to be on order of the magnetic field, and thus non-negligible. We start with the same relativistic fluid equations as \cite{Yang_2019}, derived for pair plasma by Koide \cite{Koide9}, but include shear flows for our numerical analysis, similar to the non-relativistic equation of Faganello et al.\cite{faganello}.

While it is not our primary focus, we must also mention the Kelvin-Hemlholtz instability (KHI) which forms at large, usually super-Alfv\'enic, shear flows and can induce reconnection between vorticies \cite{Chen_97, nakamura_11, Farganello_21, hamlinthesis, miura_82}. This was observed but not discussed in \cite{peery}. It is expected in the non-relativistic (both resistive and inertial) regimes that the linear tearing instability and linear KHI will both be unstable in overlapping regimes and may couple \cite{Chen_97, chenmorrison, roytershteyn, Jain_Buhner_2015}. Therefore we expect our model to recover growth of both instabilities.

The linear phase of KHI has been well studied in the non-relativistic and special relativistic regimes. Using a discontinuous velocity profile, analytical scalings have been derived \cite{peyrichon, chow23,osmanov}. Numerical methods have been used to include a finite shear width as well \cite{choudhury, berlok, ONG19721, GUDKOV, miura_82, hamlin_13, ferrari}.
Regardless of regime, it is expected that $V_{sh}>V_{A, sh}$, the  Alfv\'en velocity component parallel to the shear flow, is required for instability. However, most studies of magnetized KHI only consider a uniform magnetic field. Here we are interested in KHI where it forms at a current sheet and can affect the onset of reconnection.

In this paper we seek to characterize the onset behavior of relativistic magnetic reconnection in a shear flow, and transition to KHI, via comparison of kinetic simulation results to the linear growth rates of the tearing instability. 
In the first section of this paper, we derive a numerical solver for the growth rate and eigenvectors of the linear modes, including both tearing instability and KHI. In section 2, we show scaling of solver results. In section 3, we lay out simulation setup and overview of the growth behavior of the system. In section 4, we compare the linear solver to simulation results, and in section 5, we give some discussion and summary of our findings.\\

\section{Numerical Solver Setup}
In this section, we derive a numerical solver for the growth rates of the linear tearing and Kelvin Helmholtz instabilities. In the following, the quantity $V_{sh}$ represents the shear flow speed (normalized to $c$) and $B_g$, the guide field strength (normalized to $B_{x0}$, the reconnecting field). We use a coordinate system where $\hat{x}$ is the direction parallel to the reconnecting field, $\hat{z}$ is the direction perpendicular to the current sheet (direction of inflow) and $\hat{y}$ is out-of-plane.
As in previous work \cite{faganello, chenmorrison, yang17}, we find a system of two equations that can be solved for the perturbed variables $\tilde{\psi}$ and $\tilde{\phi}$, the flux and stream functions, assuming that these have the standard form of a Fourier mode and are small. We take the $\hat{y}$ component of Ohm's law and the $\hat{y}$ component of the curl of the momentum equation. In the end, these may be written in the form of a generalized eigenproblem and be solved numerically. 

We start with the full equations presented in \cite{Koide9, Yang_2019}.
These are reduced single fluid equations for a pair plasma that include inertial and resistive terms in Ohms law, to best compare to the kinetic simulation.

\beq
\partial_\nu (\rho U^\nu)=0,
\eeq
\beq
\label{eq1}
\partial_\nu \big{[}h\big{(}U^\nu U^\mu + \tfrac{1}{4q^4}J^\nu J^\mu\big{)} \big{]}= -\partial^\mu P+ J^\nu F^\mu _\nu,
\eeq
\beq
\label{eq2}
\begin{split}
&\tfrac{1}{4q} \partial_\nu \big{[} \tfrac{h}{q}(U^\mu J^\nu+ J^\mu U^\nu)\big{]}=\\
&-\tfrac{1}{2q} \partial^\mu \Delta P + U^\nu F^\mu_\nu - \eta[J^\mu+\rho_q(1+\Theta)U^\mu],
\end{split}
\eeq
and Maxwell's equations,
\beq
\nabla_\nu F^{\mu \nu}= J^\mu, \quad \nabla_\nu ^*F^{\mu \nu}= 0.
\eeq
From which we get the standard Faraday and Amp\'ere equations,
\beq
\nabla \cross \vb{E}= \partial_t \vb{B}
\eeq
\beq
\nabla \cross \vb{B}= \partial_t \vb{E}+4 \pi \vb{J},
\eeq
where Koide uses units normalized to the speed of light $c=\epsilon_0=\mu_0=1$.
We use Greek letters to indicate four dimensions and Roman letters to indicate three spatial dimensions so $\partial_\nu= [\partial_t, \nabla]$. In the above $F^\mu _\nu$ is the Maxwell tensor; $\rho$ is the proper mass density; $P$ is proper scalar pressure and $h=\rho c^2+\tfrac{\gamma_c}{\gamma_c-1}P$ is the relativistic plasma enthalpy with $\gamma_c$ the ratio of specific heats. For mildly relativistic plasma we use $\gamma_c=5/3$. $q=ne$ is the ``latent" charge density and $\rho_q= U_\nu J^\nu$ is the proper charge density; $\eta$ is the resistivity, and $\Theta$ is the thermal energy exchange rate. The four velocity is $U^\nu=\Gamma V^\nu$  and, unlike the equations of Koide, we use a simpler single fluid definition of the Lorentz factor, $\Gamma=\sqrt{1+U^2}$.

As with previous studies, it is simplest to separate the time and spatial components. The spatial components of Koide's equations give the 3D fluid equations used in our analysis. 
We also make a few simplifying assumptions at this step, before linearization is implemented. We will assume pressure and thermal exchange are negligible  ($P, \Theta \rightarrow 0$), and  take the system to be isothermal and homogeneous  ($\nabla ^\nu h \rightarrow 0$). Lastly, we assume incompressibility for comparison to other similar derivations ($\nabla \cdot \vb{U} \rightarrow 0$) \cite{faganello, Yang_2019} .

We use a normalization scheme equivalent to the units used for the simulations presented in this work. Velocity is normalized to the speed of light $V_o=c$, length scales to the electron inertial length (equivalent to the ion inertial length for pair plasma) $L_o=d_e=c/\sqrt{n_oe^2 4 \pi/m_e}$, and time to the plasma frequency $t_o=\tau_{pe}=1/\omega_{pe}$.
The background density is normalized to unity $n_o=1$ and magnetic fields are normalized to $B_o=\omega_{peo} m_e/c$, which enforces $\omega_{peo}=\omega_{ceo}$. Electric fields are normalized to $E_o=B_oc$. Additionally, we will write $\frac{h}{4 n^2}= h_J$, as this term shows up in relation to the current terms in Eqs. (\ref{eq1}) and (\ref{eq2}). 

After implementing these normalizations, we arrive at the two equations used in our numerical analysis. The momentum equation;
\beq
\label{momstart}
\begin{split}
 h \partial_\nu (U^\nu \vb{U})+h_J \partial_\nu (J^\nu \vb{J}) \\
= (\nabla \times \vb{B}- \partial_t \vb{E}) \times \vb{B} + \rho_q \Gamma \vb{E}
\end{split}
\eeq
and Ohms law;
\beq
\label{omstart}
\begin{split}
\Gamma \vb{E}= -\vb{U} \times \vb{B}+ S^{-1} (\vb{J}+ \Gamma \rho_q \vb{U})]\\
+ h_J \partial_\nu (J^\nu \vb{U}+ U^\nu \vb{J}).
\end{split}
\eeq

We consider a force-free equilibrium with a shear flow parallel to the reconnecting field
\beq
\vb{B}_0=B_{x0} \tanh(z/L_B)\hat{x}+B_y (z) \hat{y}
\eeq
\beq
\vb{U}_0=U_{x0}\mbox{tanh}(z/L_s) \hat{x}
\eeq
\beq
\vb{E}_0(z)= -(U_{x0}B_g/ \Gamma_0) \hat{z}
\eeq
where the functional form of $B_{y0}$ is determined by the force-free condition
\beq
B_{y0}(z)^2+ B_{x0}(z)^2-E_z(z)^2=B_{x0}^2+B_g^2-(B_gU_{x0})^2
\eeq
The initial current and charge separation are derived from these quantities. The charge is assumed to be split evenly between ions and electrons so the single fluid density remains uniform.
Unless otherwise stated, we set $L_s=1/2 L_B$ for all tearing calculations, as well as simulation runs. We assume two dimensionality; the equilibrium values only depend on $z$ and no values will be allowed to vary in the out of plane direction $\hat{y}$.

Linearization is done in the standard way \cite{Yang_2019, Betar, ofman, Tolman, baalrud, ali}, for perturbed quantities $\tilde{\psi}$ (the flux function) and $\tilde{\phi}$ (the stream function) denoted by a tilde and defined as,
\beq
\tilde{\vb{B}}= \hat{y} \times \nabla \tilde{\psi} =(\partial_z \tilde{\psi}) \hat{x} - (\partial_x \tilde{\psi}) \hat{z}
\eeq
\beq
\tilde{\vb{U}}= \hat{y} \times \nabla \tilde{\phi}= (\partial_z \tilde{\phi}) \hat{x} - (\partial_x \tilde{\phi} )\hat{z}
\eeq
with the form
\beq
\tilde{f}=\tilde{f}(z) e^{i kx+\gamma t}
\eeq
which gives, by definition, $\partial_t \rightarrow \gamma$ and $\partial_x \rightarrow ik$, and for all quantities $\partial_y \rightarrow 0$. For simplicity we will also write $\partial_z f \rightarrow f'$.
We may then find the linear quantities that depend on $\tilde{\phi}$ and $\tilde{\psi}$. 

Unlike in previous work, we will not drop the electric field terms. $E_z$ and $E_x$ are given by the ideal Ohms law and $E_y$ is given by Faraday's law. The current is given by Amp\'eres law. 
This results in derived perturbation quantities in the form $ \vb{E}= (\tilde{E}_x, \tilde{E}_y, E_{z0}+\tilde{E}_{z0})$ and $\vb{J}=(\tilde{J}_x, J_{y0}+ \tilde{J}_y, \tilde{J}_z)$, with components given in  appendix A for the main solver.

The use of Faraday's law and inclusion of the displacement current results in a nonlinear term in the time derivative, therefore we also make the assumption that $\gamma$ is small ($\partial_t^2 \rightarrow0$). Using these definitions, we then derive the linear forms of Eqs. (\ref{momstart}) and (\ref{omstart}). 

The resulting equations may be discretized into a generalized Eigenvalue problem,
\beq
\gamma \begin{pmatrix}
A_1 & A_2 \\
A_3 & A_4 \\
\end{pmatrix}
\begin{pmatrix}
\tilde{\psi}\\
\tilde{\phi} \\
\end{pmatrix}=
\begin{pmatrix}
B_1 & B_2\\
B_3 & B_4 \\
\end{pmatrix}
\begin{pmatrix}
\tilde{\psi}\\
\tilde{\phi}\\
\end{pmatrix}
\eeq
with coefficients given in Appendix A. 
This can then be solved numerically for a range of parameters.

\begin{figure}[th]
  \centering
  \includegraphics[width=\linewidth]{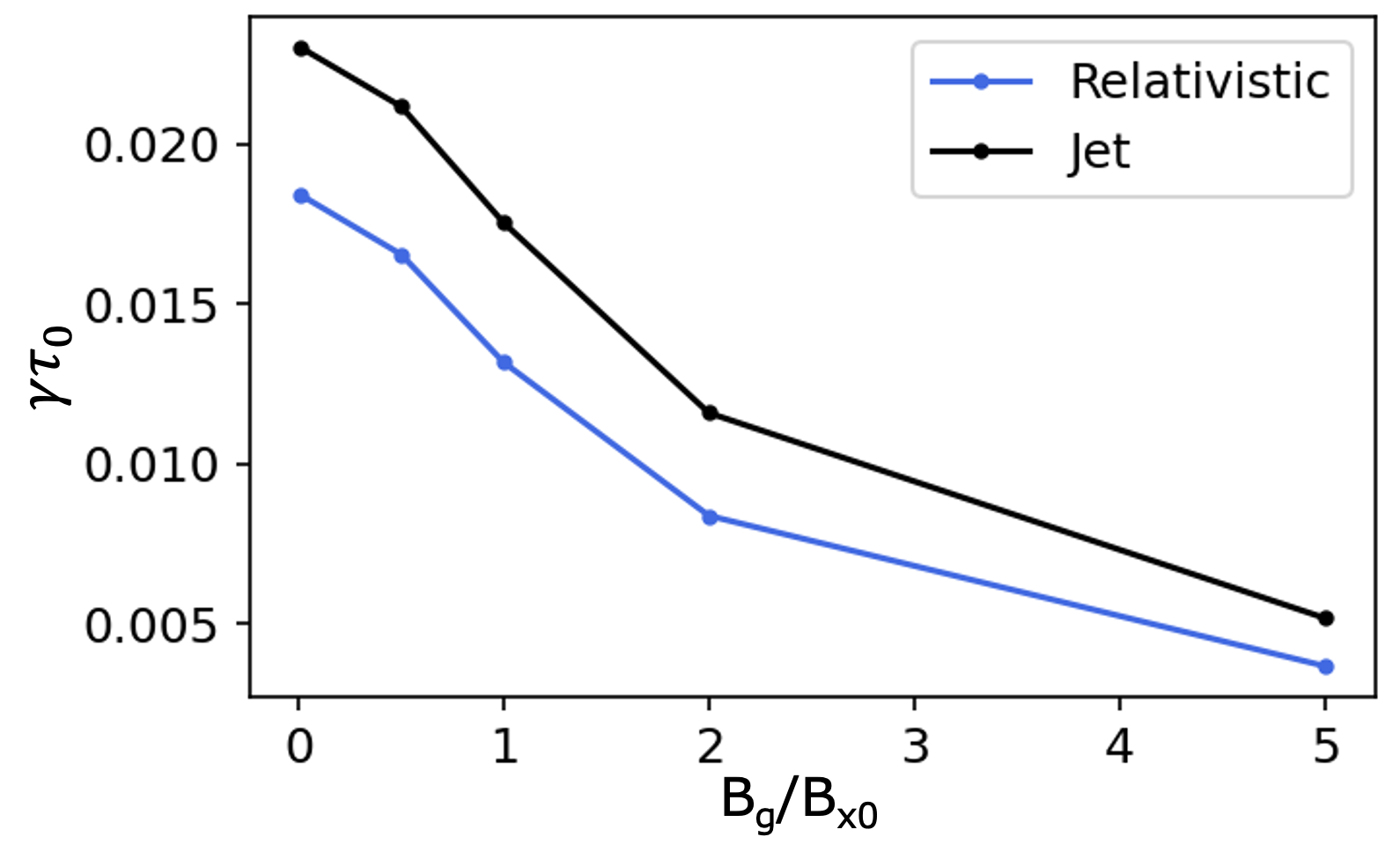}
  \caption{The scaling of the tearing growth rate with guide field strength from the relativistic solver compared to \textit{Jet}, the version of the solver using two-fluid contributions in the Lorentz factor, in the inertial regime.}  
  \label{jet} 
\end{figure}

Lastly we discuss the Lorentz factor. The linearized form of $U^0$ used in the main solver becomes
\beq
\label{gam}
\tilde{U}^0=\tilde{\Gamma}= \tilde{\sqrt{1+U^2}}=(1/\Gamma_0) U_{x0}\tilde{U}_x
\eeq
found by taking the Taylor series with respect to the single fluid velocity. However, the equations of \cite{Koide9} define the single fluid velocity as $U^\mu=(n_e U^\mu_e + n_i U_i^\mu)/n$ which gives a full Lorentz factor
\beq 
\label{full}
U^0=\left(n_e \sqrt{1+U^2_e} + n_i \sqrt{1+U^2_i} \right)/n,
\eeq
which includes contributions from the current that do not appear in the definition used in Eq. (\ref{gam}). For a force free current sheet, we have both $J_x$ and $J_y$, but $J_y$ will be the main contribution. 
Due to the complexity  eq \ref{full} would add, it is not implemented in the main solver. However, results are validated through a comparison to a version of the solver which uses Eq. \ref{full} in the limit that $V_{sh}=0$. We refer to this version of the solver as $Jet$. Details are given in Appendix B.

Results from $Jet$ and the main solver are compared in Fig. \ref{jet}, for five values of guide field.  We can see that the prediction for the growth rate from $Jet$ (black line) is similar to that of the main relativistic solver (blue line). 
The magnitudes of the growth rate remain close for the range of guide field values, which gives us confidence that the approximation we use for $\Gamma$ is, in general, valid for our system. We present some further benchmarking of the code in Appendix C, comparing the solver to familiar limits in literature.

\section {Scaling with Shear Flow}
\begin{figure}[ht]
  \centering
  \includegraphics[width=\linewidth]{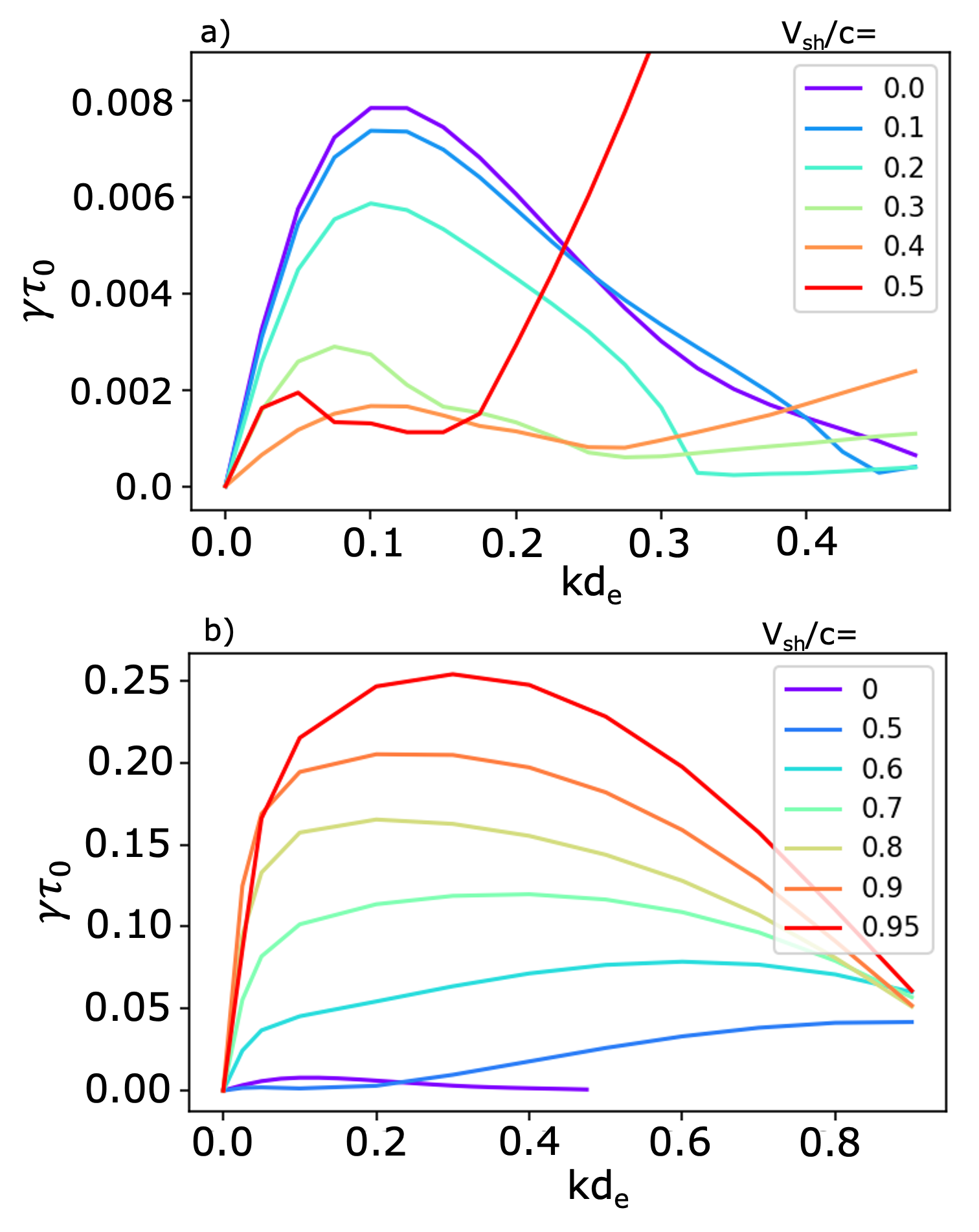}
  \caption{Linear growth rates for $\sigma_{x0}=100$,  $B_g=B_{x0}$ as a function of k. Panel a) shows growth rates for $V_{sh} \leq 0.5c$, panel b) shows growth rates for $V_{sh} \geq 0.5c$.} 
  \label{later} 
\end{figure}

In this section, we further characterize the linear growth with varying shear flow speed, in the relativistic regime of interest.
Fig. \ref{later} shows the dispersion relation for $B_g=B_{x0}$ for the full range of shear flows available. Fig. \ref{later}a shows the growth rates of tearing instability at lower $V_{sh}$ and $k$ and Fig. \ref{later}b) shows the growth rates of Kelvin Helmholtz instability at larger $V_{sh}$ and $k$. We also plot the dispersion curve for $V_{sh}=0$ in Fig. \ref{later}b for better comparison. 

The linear tearing has a maximum growth rate of $\gamma \tau_{pe}=0.008$  at $kd_e=0.1$, for the case with no shear flow, and a decreasing maximum growth rate as shear flow increases. For $V_{sh}>0.3c$ there is a clear transition in the behavior, and above $V_{sh}>0.5c$, the Kelvin Helmholtz modes become unstable. 
At $V_{sh} \sim V_{A,rec}$ (for $B_g=B_{x0}$ this is $V_{A,rec} \simeq 0.7c$) the growth rate of KHI is an order of magnitude larger than that of tearing, and at highly relativistic flows, it can be two orders of magnitude larger. Predictions for $V_{sh}>0.95c$ are not shown due to a loss of validity of the solver in this regime, due to limitations of the fluid model, as discussed in Appendix C. The wavenumber of the largest KHI mode increases as we would expect, both because, for relativistic KHI larger wavenumbers remain unstable \cite{hamlin_13}, and because we use $L_s=0.5L_B$, causing the characteristic length of KHI to be half that of tearing modes. This helps distinguish the mode transition.

For $V_{sh} \simeq 0.4c$ the growth rate behavior does not resemble either pure tearing or pure KHI modes, but is clearly related to the presence of both the shear flows and the current sheet, since it does not appear in a uniform magnetic field (see Fig. \ref{khis} in Appendix C).
We believe that this is an intermediate regime arising from the coupling of tearing and KHI in the linear phase. Note, this is a distinct behavior from vortex induced reconnection, which occurs in the nonlinear stages of KHI and therefore cannot be captured by the linear solver. Several previous studies have also observed regimes where KHI and tearing exist without either dominating \cite{chenmorrison, voslion, grasso}. 

\begin{figure}[h!]
  \centering
  \includegraphics[width=0.9\linewidth]{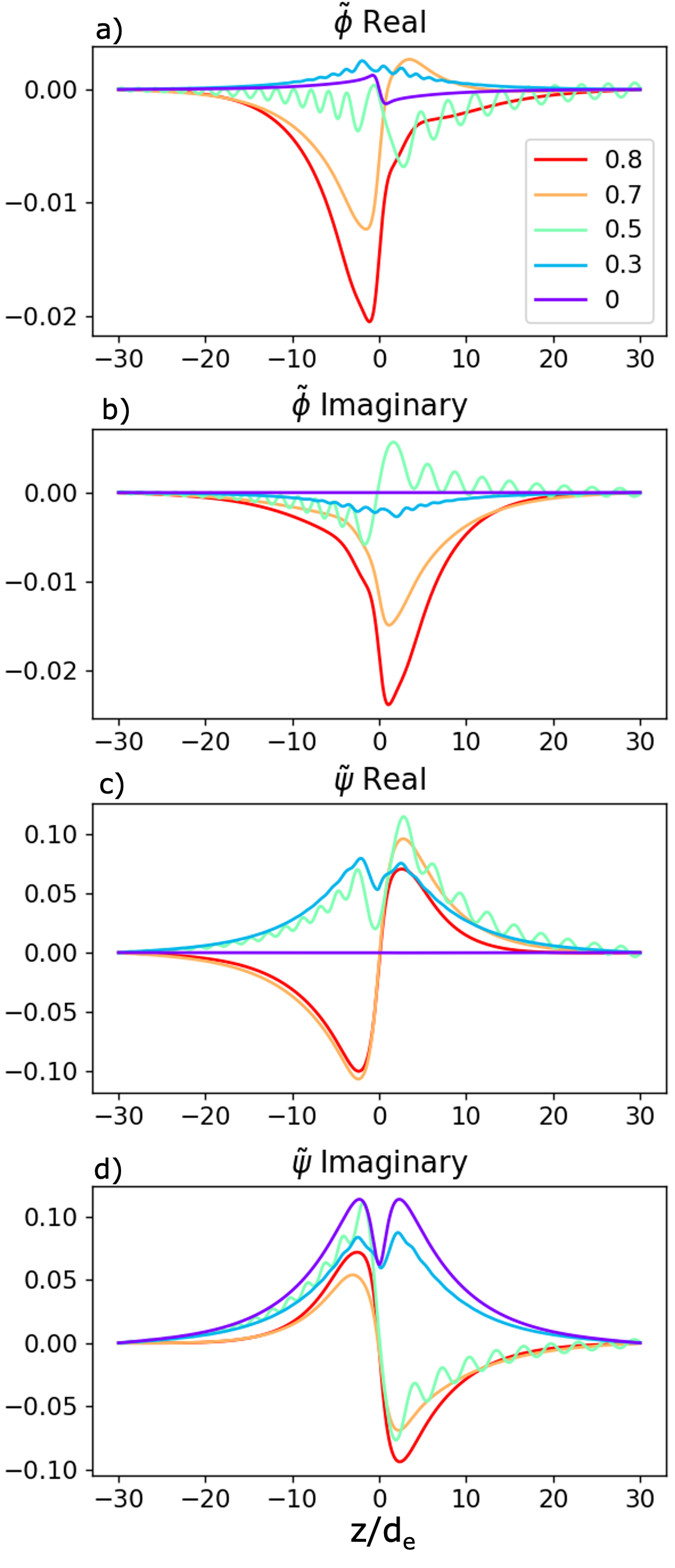}
  \caption{The real and imaginary parts of the perturbed eigenvectors for flux functions and stream functions for $V_{sh}$= 0, 0.3c, 0.5c, 0.7c, 0.8c at k$d_e$=0.1, recovering the standard tearing eigenfunctions for $V_{sh}=0$ and showing the transition to KHI for $V_{sh} >$0.5c}
  \label{eigens} 
\end{figure}

To better understand this behavior, we plot the eigenvector profiles of $\tilde{\psi}$ and $\tilde{\phi}$ as a function of $z$, for four values of shear flow (Fig. \ref{eigens}). For the case with no shear flow, best seen in Fig. \ref{eigens}d, we recover the expected two-peaked structure that is characteristic of tearing modes \cite{jonghe, Betar, ottovani, Demidov_2025, gourgouliatos, Demidov_Lyubarsky, einaudi, ottovani, zanna}. For the case with $V_{sh}=0.3c$, which has a significantly lowered growth rate, we see only slight changes in $\tilde{\psi}$, specifically the two peaked structure is still present, but $\tilde{\phi}$ has become distorted. 
When $V_{sh}=0.5c$, the eigenfunctions are even more distorted, displaying characteristics of both tearing and KHI. 
Lastly there is a transition in behavior at $V_{sh}=0.7c$ (and the imaginary component of $\tilde{\psi}$ for $V_{sh}=0.5c$) and the two peaked structure no longer appears, as KHI becomes the dominant mode. Several studies \cite{einaudi, Jain_Buhner_2015} recover a similar transition from tearing to KH modes in the non-relativistic regime. 

We are clearly seeing four regions of behavior; standard tearing at $V_{sh}=0$; distorted but still unstable tearing at low $V_{sh}$; an intermediate region; and KHI at high $V_{sh}$. This behavior is also recovered at other guide field strengths, though the range of shearflow that defines each region shifts lower with the in-plane Alfv\'e velocity.
We also see that the eigenfunctions with $V_{sh}>0$ can be slightly asymmetric (similar to some results found for double tearing modes \cite{voslion}). However, in this case we believe it to be due to the vertical motional electric field $E_z$.

\begin{figure*}[ht]
  \centering
  \includegraphics[width=\textwidth]{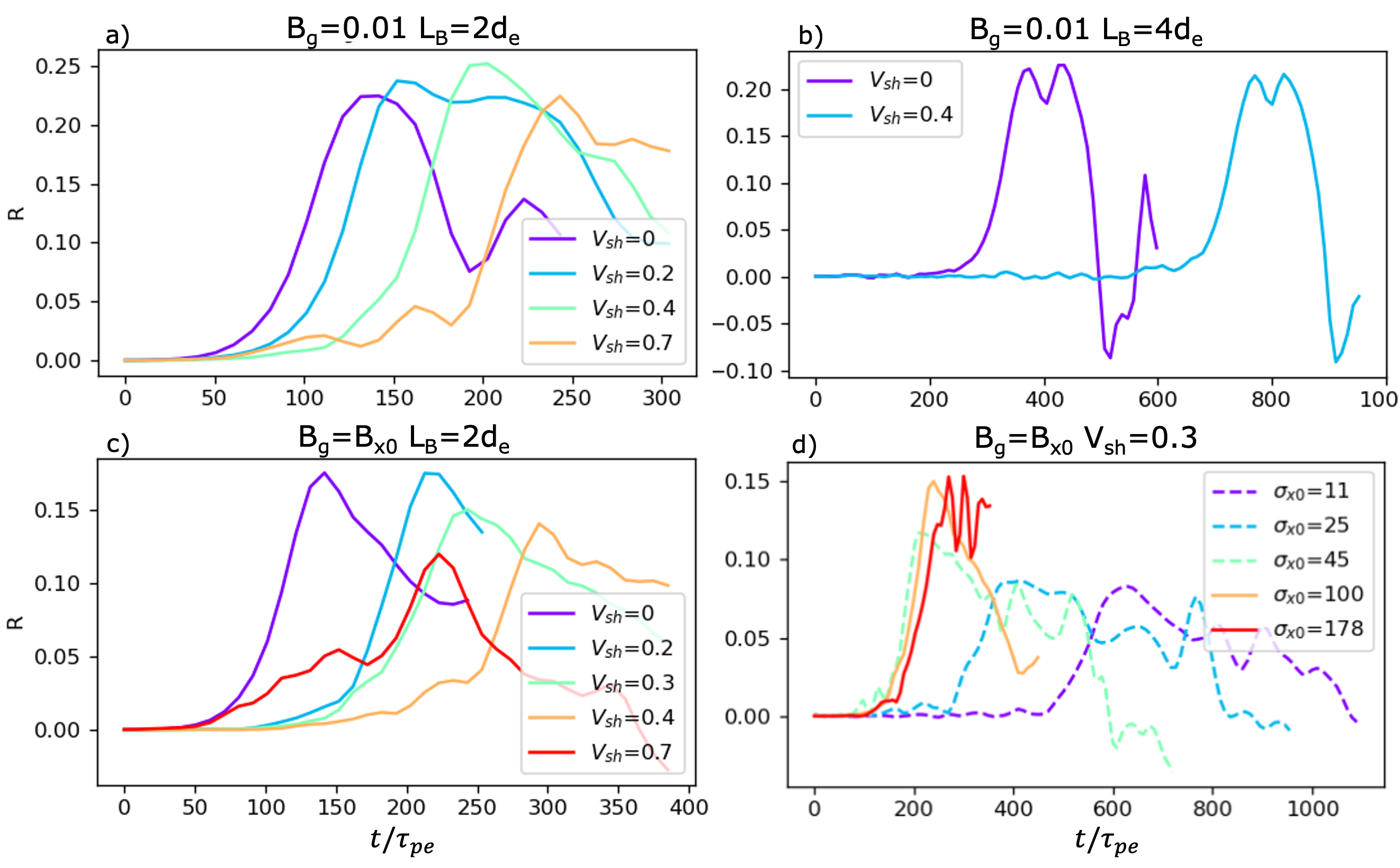}
  \caption{Reconnection rates for simulations with $B_g/B_{x0}=$0.01 and 1 and shear widths $L_B=2, 4 d_e$. Panels a-c) show runs with $\sigma_{x0}=100$ and varying $V_{sh}$, normalized to $c$. Panel d) shows simulations with $B_g=1$, $V_{sh}=0.3c$ and varying  $\sigma_{x0}$, solid lines show simulation in Fig. \ref{table}, dashed lines indicate the smaller simulation with lower $\sigma_{x0}$.} 
  \label{rat} 
\end{figure*}

\begin{figure*}[ht]
  \centering
  \includegraphics[width=\textwidth]{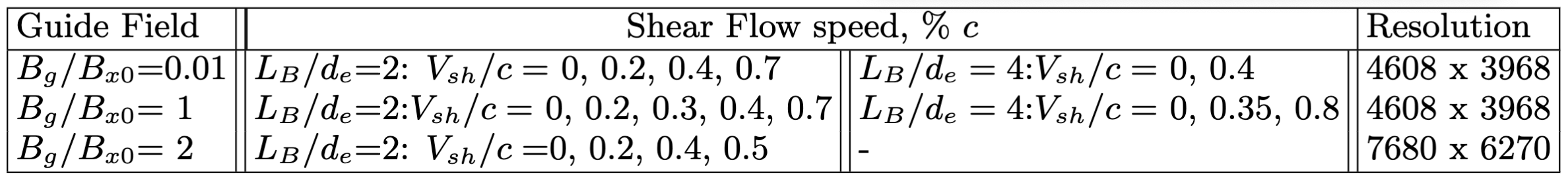}
  \caption{The set of main simulations used, guide field is normalized to $B_{x0}$, sheet width $L_B$ to $d_e$, shear flow speed is normalized to $c$ and resolution gives number of grid points for simulation size $L_x \times L_z=130d_e \times 150d_e$} 
  \label{table} 
\end{figure*}


\section{Simulation Overview}
Now we discuss simulation results, which are compared to the solver. In this paper, we use simulations with initial setup similar to that of \cite{peery}. We use a relativistic force free current sheet for pair plasma, with additional in-plane shear flows and out-of-plane guide field. In \cite{peery} , a wide current sheet with an initial central perturbation was used. In this work, we start with thin current sheets of $L_B=2d_e$ and $4d_e$ with initial perturbation, which allows tearing instability to begin spontaneously in a shorter time. We examine a range of shear flow speeds from 0 to $c$,  and guide field values from $B_g/B_{x0}=$ 0.01 to 5. Most simulations are done with $B_{x0}=15$ or $\sigma_{x0}=100$. Predominantly, we focus on simulations listed in Fig. \ref{table}. These have system size $L_x \times L_z=130d_e \times 150d_e$ and resolution as listed, which increases with $B_g$ to resolve the gyro-radius with the criteria in \cite{guo_15}. 

While the numerical solver assumes a simple sinusoidal dependence on $x$ (and therefore, an effectively infinite domaine) the simulations are periodic in the $x$ direction. The boundaries in the $z$ direction are conductive for the field and reflective for particles
The wavelength of the plasmoids is small enough, especially during early growth, that development will not be affected by the box length. All simulations had an aspect ratio large enough that tearing mode development in the center was necessarily far from the box boundaries to avoid any interference. Further details about simulation setup are given in Appendix D. 

Figure \ref{rat} shows reconnection rates as a function of time for runs in Fig. \ref{table} as solid lines, and additional simulations as dotted lines. Rates are found as the rate of change in the flux converted between the primary X-point and the O-point;
\begin{equation}
R \equiv \frac{c E_y|_{\rm xline}}{B_{x0} V_{Ax0}}= \frac{1}{B_{x0} V_{Ax0}}\frac{d \Psi}{dt}
\end{equation}
where $\Psi= {\rm max}[A_y]-{\rm min}[A_y]$ along the  $B_x=0$ line, and $A_y$ is the out-of-plane component of the vector potential. In Fig. \ref{rat}d, four different values of initial reconnecting field strength $B_{x0}=$5, 10, 15, 20 are shown, corresponding to $\sigma_{x0}= 11,45,100,178$  each with $B_g/B_{x0}=1$ and $V_{sh}=0.3c$. All other plots use $\sigma_{x0}=100$. Fig. \ref{rat} shows a delay in the onset time for runs with higher shear flow speed and guide field strength, as was found in \cite{peery} (shown in Fig. \ref{one}). We find a maximum reconnection rate faster than R=0.1, most likely due to the rapid generation of plasmoids.
Unlike in \cite{peery}, which starts from a central fluctuation and does not have a linear tearing mode phase, these plasmids come from the initial phase of tearing and not a secondary collapse of the current sheet, and thus occur even with a guide field.

Unsurprisingly, the shear width also has a strong impact on the onset time and evolution. The thin current sheets $L_B=2d_e$ are very unstable, and fast reconnection starts almost immediately. In the $L_B=4 d_e$ runs, the delay is much longer, consistent with previous observations that linear growth decreased with sheet width \cite{betar20, loureiro, honshino21}.
Lastly, we examine the effects of changing the initial magnetization. We would expect the linear growth rate and the reconnection rate to be strongly affected by upstream magnetization (note that $R$ is still normalized to the reconnecting field). We find that this is only the case for lower $B_{x0}$. For $B_{x0} \geq 10$, the onset time appears to be very similar.  The same trend is recovered in the numerical solver, so we can conclude that for a large enough initial magnetic field, the growth rate converges.

Most simulations start from a thin current sheet where a chain of plasmoids form with multiple X-points, which grow until one X-point dominates. 
The exception are the runs at high shear flow, most notably $B_g/B_{x0}=1, V_{sh}=$0.4c shown in Fig. \ref{rat}d, for which the initial instability that disrupts the current sheet is no longer tearing. Instead, an intermediate behavior between KHI and tearing is shown. Note this is not VIR; reconnection is still the dominant nonlinear behavior.

\begin{figure*}[ht]
  \centering
  \includegraphics[width=0.9\linewidth]{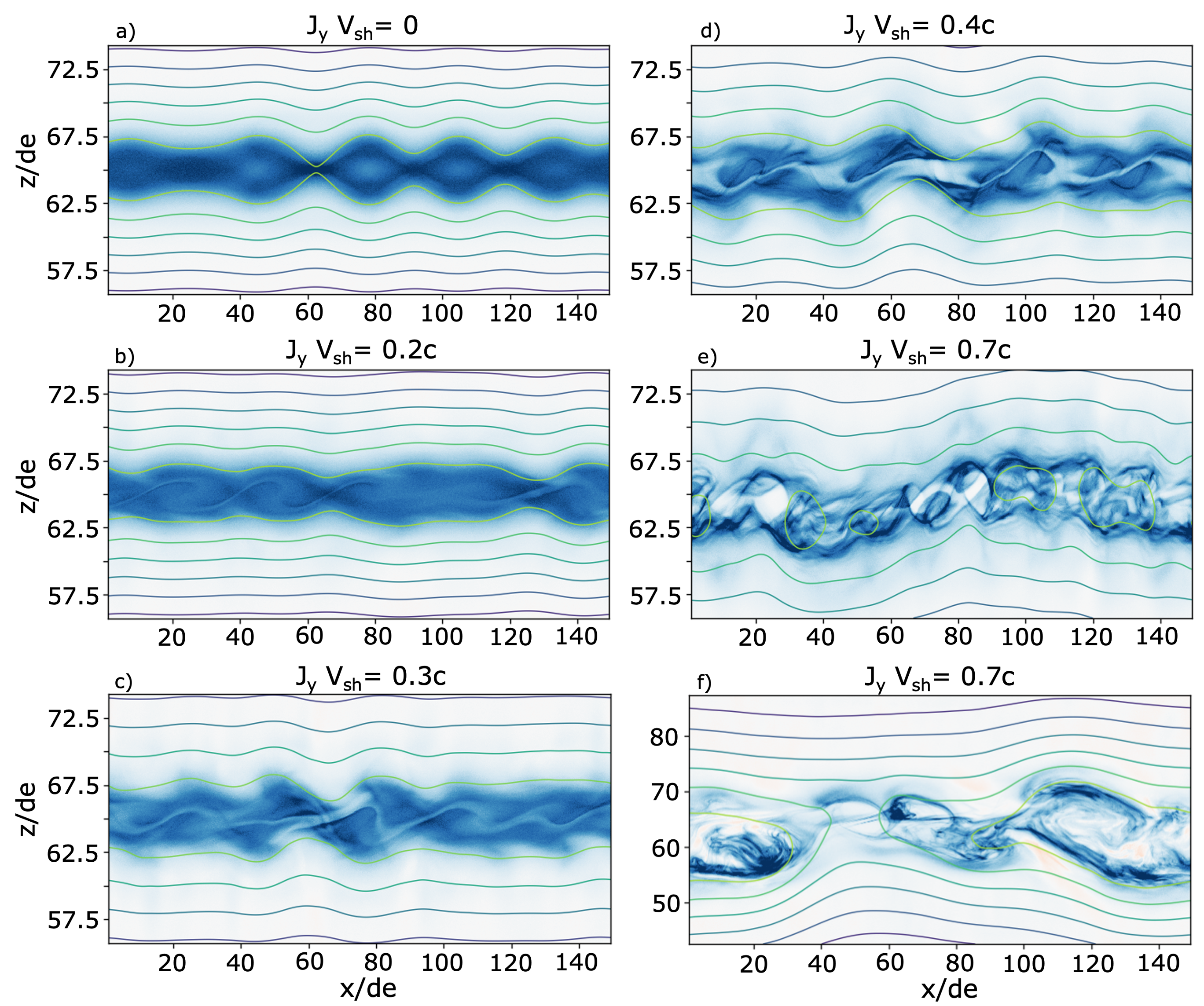} 
  \caption{Panels a - e show snapshots of $J_y$ for $B_g/B_{x0}=1$, $L_B=2d_e$ and $V_{sh}$= 0c, 0.2c, 0.3c, 0.4c, 0.7c respectively, to demonstrate the transition behavior at increasing shear flow. Panel f shows the same simulation as panel e), at a later time to show vortex formation. The plots are taken at $t/\tau_{pe}$= 120,180, 200, 250, 150, 250 respectively.}
  \label{trak} 
\end{figure*}
To examine the transition as shear flow increases, Fig. \ref{trak} shows snapshots of the out-of-plane current and magnetic flux lines from simulations with $B_g/B_{x0}=1$, $L_{B}=2 d_e$ for $V_{sh}=$0, 0.2c, 0.3c, 0.4c and 0.7c, chosen to be just after the initial onset of reconnection so the plasmoid formation is large enough to be identifiable. Fig. \ref{trak}f shows a time during the later nonlinear phase of the run in Fig. \ref{trak}e, to show the formation of a KH vortex. 

Fig. \ref{trak}a shows undisturbed tearing modes and a chain of plasmoids form, Fig. \ref{trak}b also shows plasmoid formation but with some vortex-like distortion in the current sheet, which is greatly enhanced in Fig. \ref{trak}c. As plasmoids form here, they are also rotated, but the dominant behavior is still tearing.
In Fig. \ref{trak}d the current sheet is very distorted. While the run evolves to have a fast reconnection phase, as Fig. \ref{rat} shows, it is not clear that this is due to the standard tearing evolution. KHI may not be the dominant instability, but we do see the beginnings of vortex formation. We identify this with the transition regime where neither tearing instability or KHI is dominant, but characteristics of both are present. Fig. \ref{trak}e, in comparison, shows a run that is Kelvin-Helmholtz unstable.
Small wavelength vorticies form first (in Fig. \ref{trak}e) and then at later times the whole current sheet is disrupted and large scale vorticies form (in Fig. \ref{trak}f), so that unlike in runs with lower shear the reconnection recorded is vortex driven.
This transition happens for other guide field strengths as well, but the formation of KHI vorticies depends on the initial conditions. In lower guide fields, even for shear flows above the Alfv\'en velocity, the current sheet becomes more turbulent. While in high guide fields, and for wider current sheets, large scale vorticies can form.

We will also note that other kinetic instabilities may develop in the simulation \cite{yoon}, which would not be captured by the fluid solver.
So we must be careful when we say reconnection is suppressed, as 
there are cases where it may be accurate to say that the tearing instability is suppressed, but reconnection will still occur.

\section{Comparison to the Linear Solver}
\begin{figure}[h!]
  \centering
  \includegraphics[width=\linewidth]{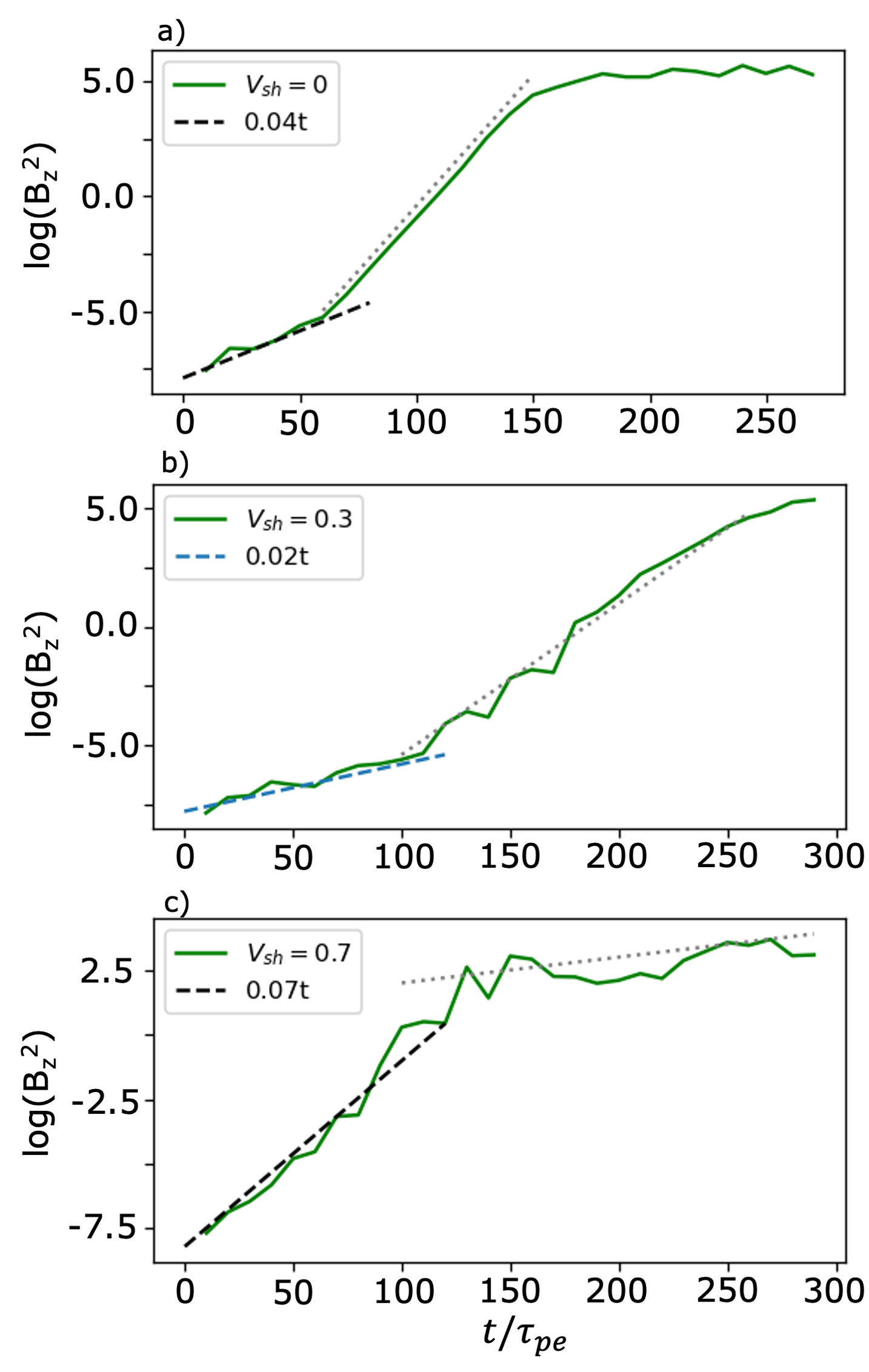}
  \caption{Time evolution of ln($B_z^2$) in simulations with $B_g/B_{x0}=1$. Panel a) shows tearing with no shear flow, panel b) shows tearing slowed by shear flow and panel c) shows KHI growth. Twice the linear growth rate is plotted with a dashed line and twice the second phase growth (fast growth for tearing) is plotted in dotted gray. $V_{sh}$ is normalized to $c$ and $t$ to $\tau_{pe}$.}
  \label{bzs} 
\end{figure}
Finally, we may look directly at the evolution of the perturbed magnetic field, used to find the linear growth rates, which we compare to results of the numerical solver. For runs with $L_B=2d_e$, a growth rate can be found directly from the time evolution of $B_z$ perturbations.
Fig. \ref{bzs} shows plots of the maximum value of log($B_z^2$) along the neutral line, for three values of shear flow, $V_{sh}=$0, 0.3c, and 0.7c with $B_g=B_{x0}$ and $L_B=2d_e$. We plot the magnetic energy to better identify the signal. Similar to what was observed by Schoeffler \cite{Schoeffler} for relativistic tearing modes, in Fig. \ref{bzs}a and b, there are two main growing phases. The first is the early linear growth phase, which happens before reconnection starts and lasts until the magnetic energy is $B^2 \sim e^{-5}$. The second has much faster growth, which was previously identified with the nonlinear growth. 
This phase extends into the fast reconnection. In Fig. \ref{bzs}c, we also show evolution of a run that is Kelvin Helmholtz unstable, which instead has an extended first linear growth phase and much slower second phase (which is consistent with the evolution of the reconnection rate in Fig. \ref{rat}, where reconnection is induced quickly, but then proceeds much more slowly to a maximum rate). Other guide field values with a thin current sheet show very similar behavior for both tearing and KHI growth.

For the lower resolution simulations and the wider current sheet runs, there is too much noise to recover a distinguishable linear phase, as also seen by \cite{Schoeffler}. And there is no linear phase in the simulations of \cite{peery} which start with a perturbation.
In these cases, we estimate an average growth rate by assuming it will scale inversely with the onset time for fast reconnection. The onset time is found as the time where $R>0.01$ across all runs, which is roughly 5-10\% of the maximum rate. This value is chosen to allow for standardization across different sets of parameters. As a result, it may not coincide precisely with the end of the linear phase, as the transition out of the linear phase can happen slightly before fast onset. However, we believe it gives an fair measure of the average growth rate of the linear phase. To compare to simulations, the linear growth rates from the solver are also normalized to $\tau_{pe}= 1/\omega_{pe}$. 

\begin{figure}[ht]
  \centering
\includegraphics[width=\linewidth]{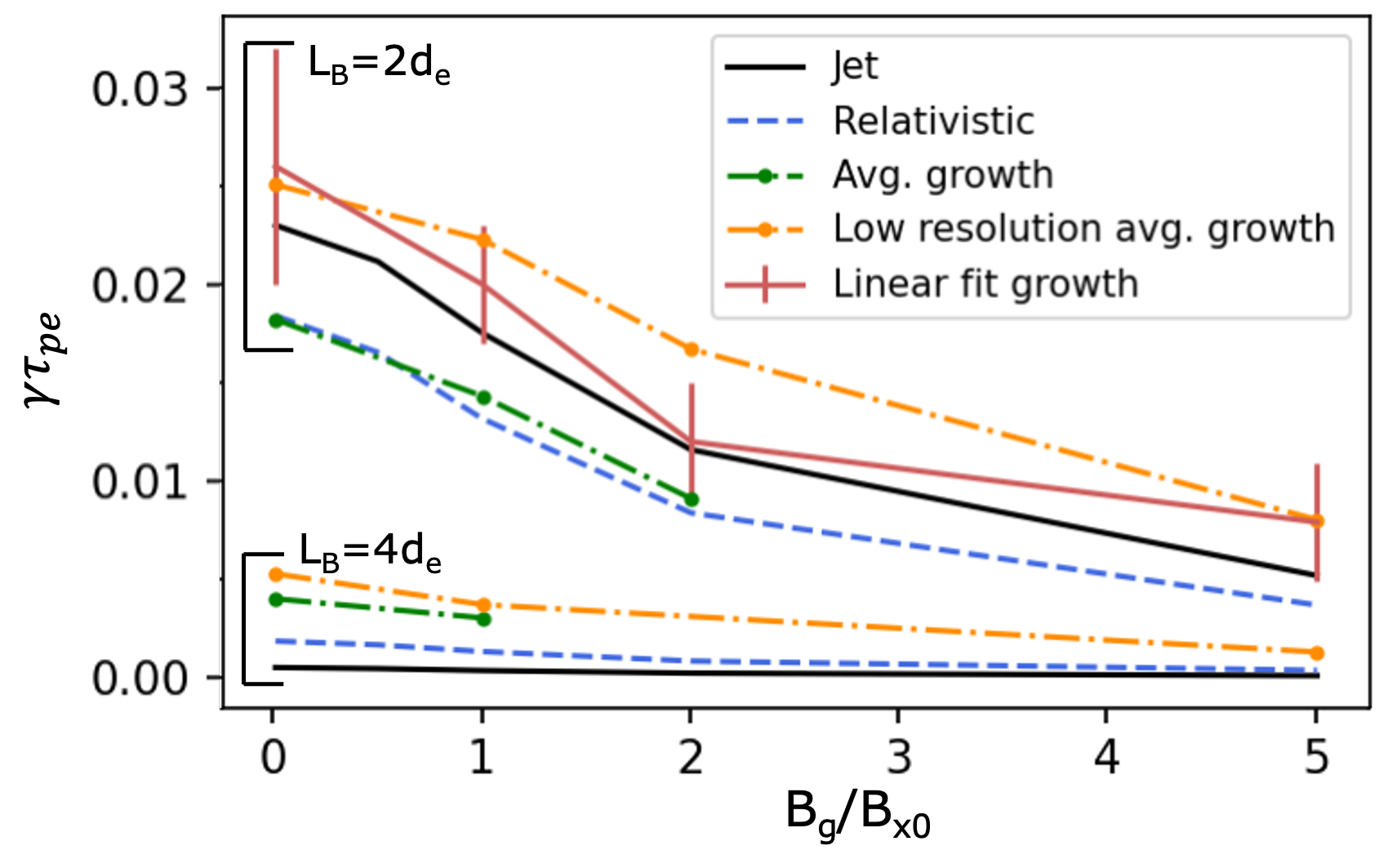}
  \caption{Growth rates with $V_{sh}$=0 as a function of $B_g$ for $L_B=2$ and $4d_e$ from the main relativistic solver (blue line) and Jet (black line), and from the simulations; growth from the linear fit found only for $L_B=2d_e$ (red line), average growth from onset time (green line) and average growth in the lower resolution simulations (orange line).} 
  \label{compar1} 
\end{figure}
  
Figure \ref{compar1} shows the growth rates from the simulation and solver calculated in several ways, with no shear flow first so we may discuss the differences in these values. Both $L_B=2d_e$ and $4d_e$ are shown. We plot results of the main relativistic solver (blue lines) and $Jet$ (black lines) which were also shown in Fig. \ref{jet}.
From the simulation, we plot growth rates found from the linear fit to log($B_z$) for $L_B=2d_e$ runs (red line) and average growth rates found from the reconnection onset  time for all runs (green lines). We also plot the average growth from several lower resolution simulations for comparison (orange lines). We have already compared $Jet$ to our main solver, and expect the main solver to be slightly underestimating the growth for a thin current sheet. The solver (blue) also underestimates the growths from the linear fit (red). The error bars here come from noise in the $B_z$ values and uncertainty from the linear fit process. The fit depends on the time period used and it is not always obvious exactly when the linear phase ends.
The average growths (green) match up well with the main solver results for a thin current sheet, indicating that these are also slightly underestimating the true growth rate, likely due to the linear growth phase ending before the onset of fast reconnection. Lastly, the average growth rates of the lower resolution runs overestimate the growth rate because reconnection onsets faster when there is more initial noise.

In the thicker current sheet, the growth rate is about an order of magnitude smaller, consistent with a scaling of $\gamma \sim L_B^{-3}$ expected by Liu et. al. \cite{ys14}. We do not plot a linear fit as the growth in this case is not well distinguished from noise. Unlike the previous case, $Jet$ predicts a smaller rate than the main solver, and both underestimate the average growth rates. This may indicate that 
tearing reaches the nonlinear phase before it is large enough to disrupt the current sheet. 
Overall we do find that, despite approximation made in the derivation, our main solver is capable of capturing the linear behavior of relativistic tearing instability to a much better extent than an order-of-magnitude estimate.

\begin{figure}[ht]
  \centering
  \includegraphics[width=\linewidth]{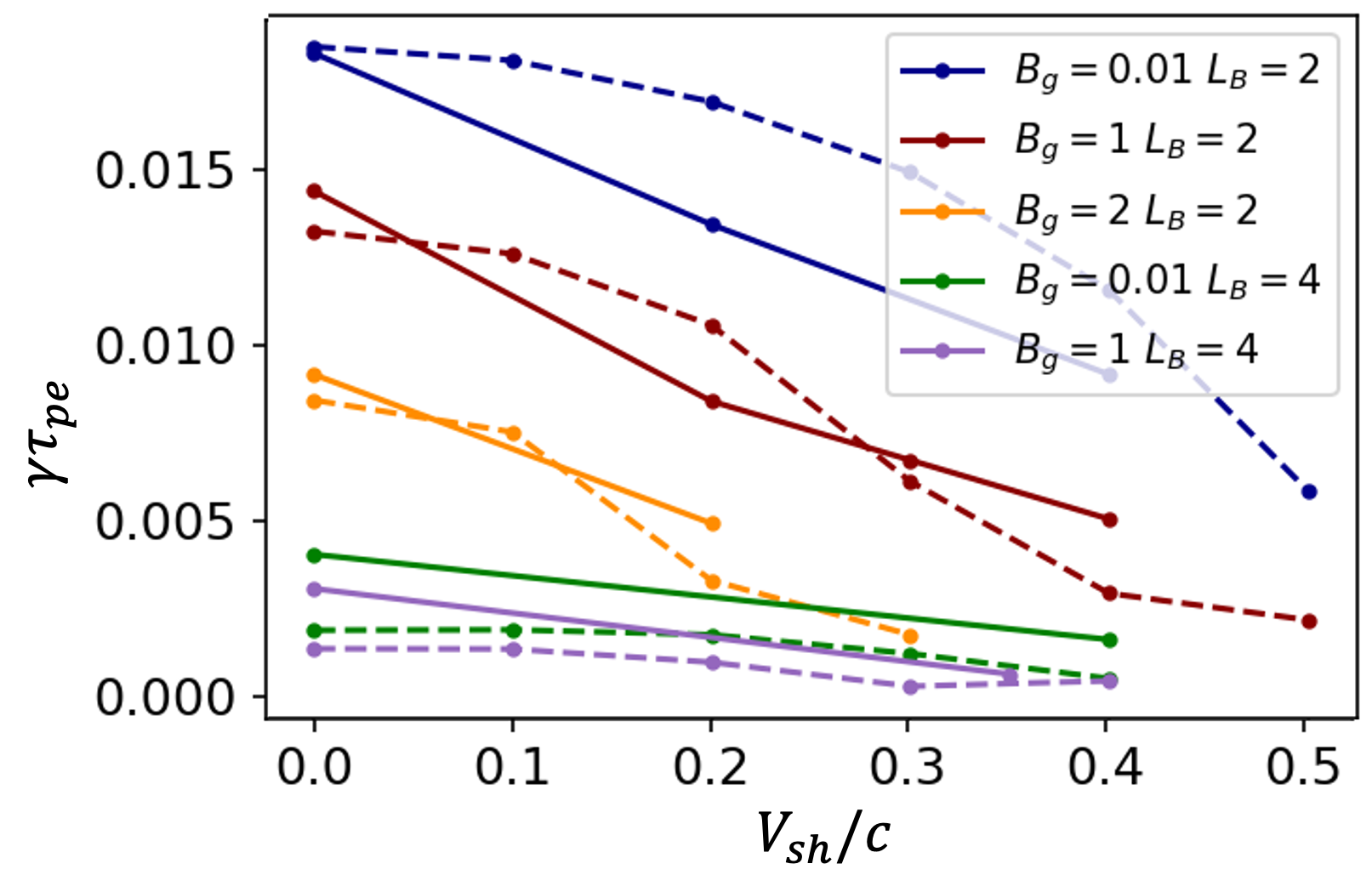}
  \caption{Linear growth rates from the solver (dashed lines) and average growth rates from the simulations (solid lines) for three values of guide field and two current sheet widths as a function of shear flow.  $B_g$ normalized to $B_{x0}$ and $L_B$ to $d_e$.}
  \label{gro}  
\end{figure}

\begin{figure}[ht]
  \centering
  \includegraphics[width=\linewidth]{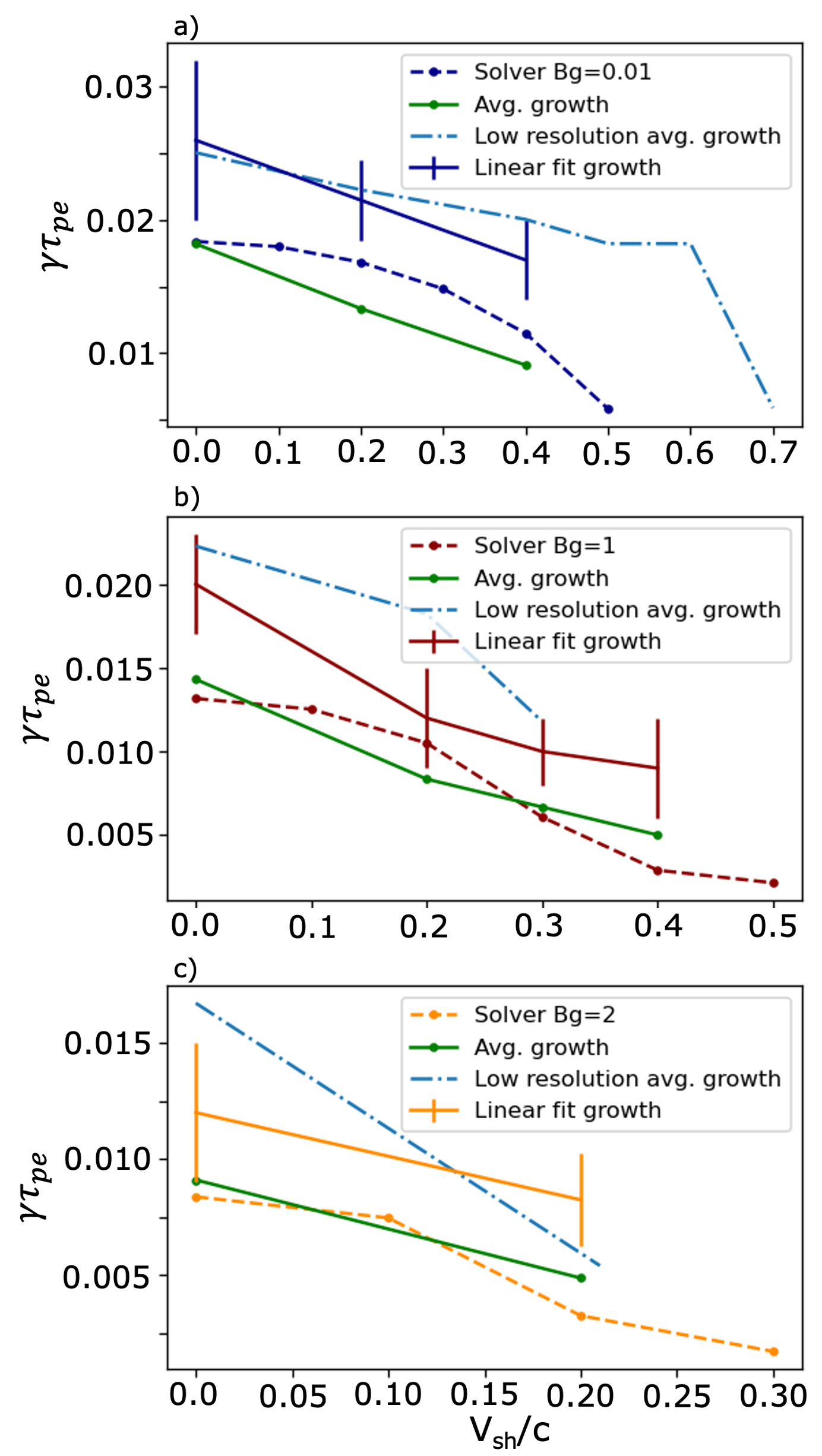}
  \caption{Growth rates as a function of $V_{sh}$ from the main solver (dashed lines), linear fits to growth from the simulation (solid lines), average growth from the simulation (green lines) and average growth from the low resolution simulations (dashed-dotted lines). Panel a) shows $B_g/B_{x0}=0.01$ panel b) shows $B_g/B_{x0}=1$, panel c) shows $B_g/B_{x0}=2$, the colors of the solver line are chosen to match those in Fig. \ref{gro}.}
  \label{comp2} 
\end{figure}

Now we examine the behavior when shear flow is also varied. Fig. \ref{gro} shows the scaling of the growth rate with shear flow speed for three values of guide field $B_g/B_{x0}=$ 0.01, 1, 2 and two current sheet widths $L_B/d_e=2$, 4.  We present them on one graph to better compare how all three parameters will affect linear growth rate and therefore only show the average growth rates. Fig \ref{comp2} shows more details on the $L_B=2d_e$ runs, for each value of guide field individually. 
We also only show shear flow values that are tearing unstable. The onset times in the case of the intermediate and KHI runs are related to the growth, as the instability is what distorts the current sheet into an X-point. But from Fig. \ref{bzs} we would not expect the evolution to be the same and thus the inverse of the onset time of reconnection does not capture KHI growth rate well.

We again find that the $L_B=2d_e$ cases match slightly better than the $L_B=4d_e$, but the same trends are still apparent across all the plotted values in both the solver and the simulation. The growth rate decreases with both increasing guide field strength and shear flow speed, as expected.
In Fig. \ref{gro} we also see that the scaling with $B_g$ and $V_{sh}$ is similar for both current sheet widths. For example, with a $B_g=B_{x0}$,
we find intermediate behavior at $V_{sh} \simeq 0.4c$ for both $L_B=2d_e$ and $4d_e$ (see also Fig. \ref{eight}), indicating that the shear flow speed at which tearing instability is no longer dominant does not depend strongly on the sheet width.
This confirms that the onset time of magnetic reconnection is closely related to the linear growth rate of tearing instability.

\begin{figure}[ht]
  \centering
  \includegraphics[width=\linewidth]{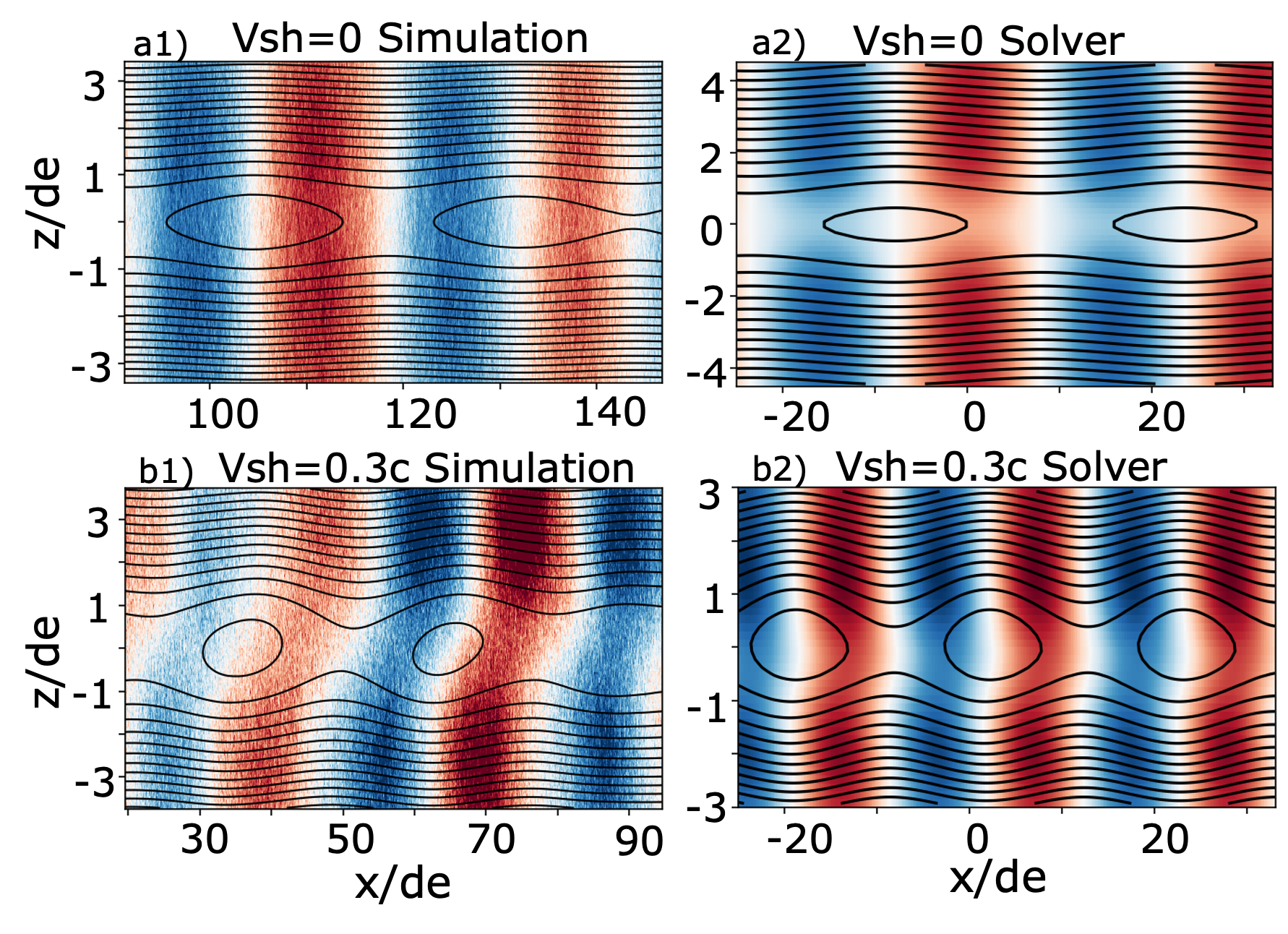}
  \caption{Panels a1) and b1) plot $B_z$ and contours of the total $\psi$ function during the linear phase of simulations with $B_g=B_{x0}$ $V_{sh}=0, 0.3c$ respectively. Panels a2) and b2) show $B_z$ and 2D contours of the total $\psi$ function found from the eigenvectors of the relativistic solver.}
  \label{eigscomp} 
\end{figure}
We may also compare the eigenvectors; Fig. \ref{eigscomp} shows 2D plots of $B_z$ and contours of the flux function from the solver, compared to early plasmoid formation in the simulations. Fig. \ref{eigscomp}a1 and  a2 show the case with $B_g=B_{x0}$ $V_{sh}=0$ and we find the standard morphology; the tearing modes grow with a width of $\sim d_e$ scale, and $\sim 15d_e$ wavelength,
and are in general symmetric. Fig. \ref{eigscomp}b1 and b2 show the case with $B_g=B_{x0}$, $V_{sh}=0.3c$ and there is a slightly distortion in the plasmoid shape and magnetic field, but at this early time it is still a subtle effect. At later times this will evolve into the deformed and tilted structure shown in Fig. \ref{trak}c.

\section{Discussion}
In this study we developed a numerical solver for the linear growth rate of relativistic tearing instability in the presence of a shear flow, to model the onset time of magnetic reconnection. As shear flow increases, the tearing instability is stabilized, and the onset of reconnection is delayed. At higher shear flow speeds, the current sheet is still disrupted by an instability intermediate between tearing and KHI. This leads to X-point formation on a similar time scale, allowing reconnection to happen at a broader range of shear flow values than predicted.

\begin{figure}[t]
  \centering
  \includegraphics[width=\linewidth]{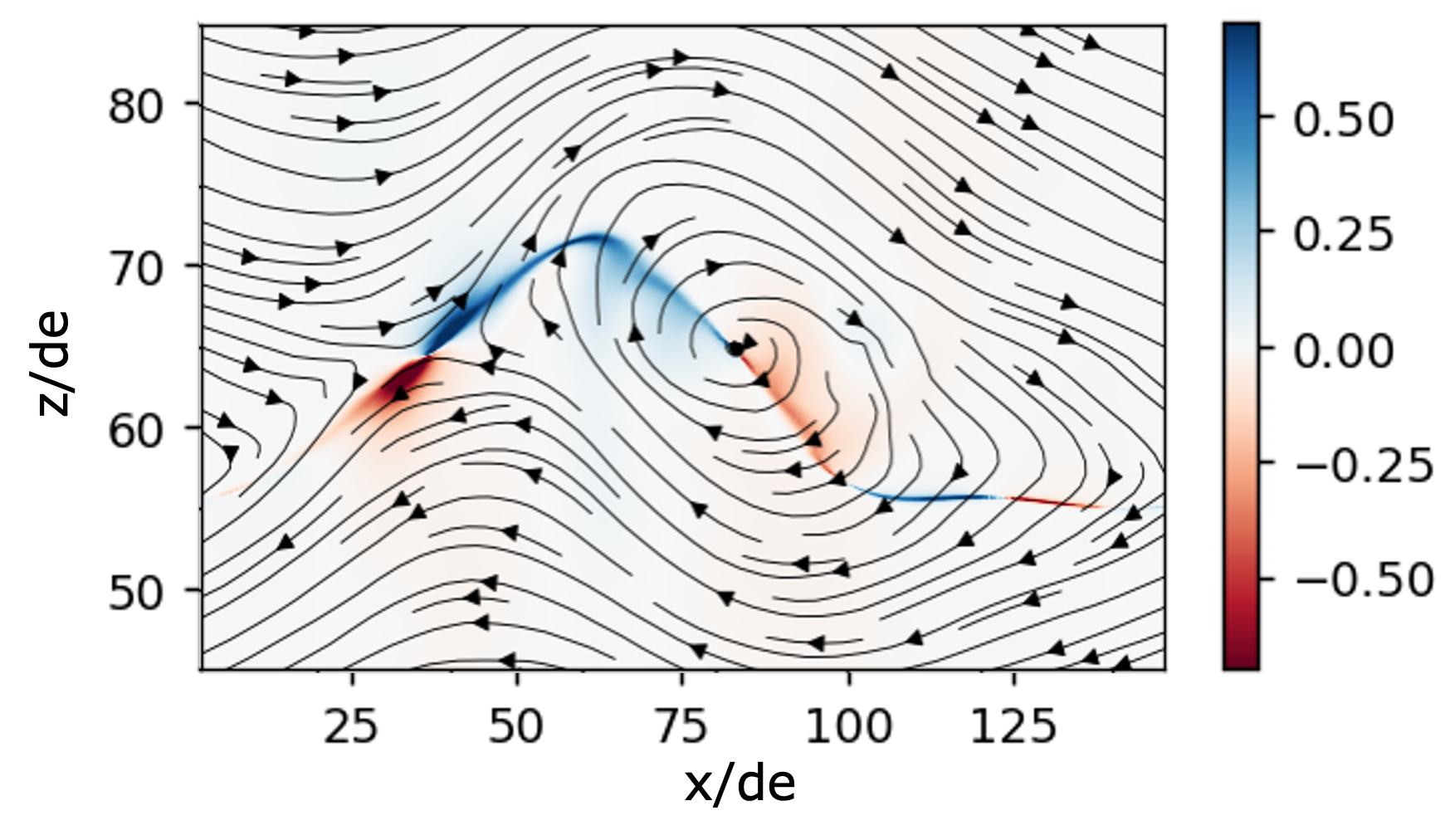}
  \caption{An example of vortex induced reconnection with $B_g=B_{x0}$, $V_{sh}=0.8c$. We plot the x component of the magnetic flux transport speed  \cite{li_21} to identify reconnection sites (see (x,z) $\simeq 40,63$), and streamlines of in-plane velocity to show the vortex formation.} 
  \label{vir} 
\end{figure}
With higher shear flows, the dominant nonlinear mode becomes KHI, which can sufficiently warp the current sheet to create an X-point. We have not examined the dynamics of this vortex induced reconnection, as the linear solver does not model the late nonlinear stages. However, we find frequently in the simulations that vortex induced reconnection (VIR) forms \cite{takuma22, nakamura_11, kevin} as shown in Fig. \ref{vir}. This indicates that reconnection can still be unstable in some form for almost the whole range of shear flow speeds, if additional onset mechanisms besides tearing instability are considered. 
This appears to contradict results of \cite{peery}, which expects the steady state reconnection jet to be suppressed at high shear flow. But the force balance used in that work was not designed to take into account highly deformed current sheets or vortex induced reconnection, which is driven by a different mechanism. 

\begin{figure}[t]
  \centering
  \includegraphics[width=\linewidth]{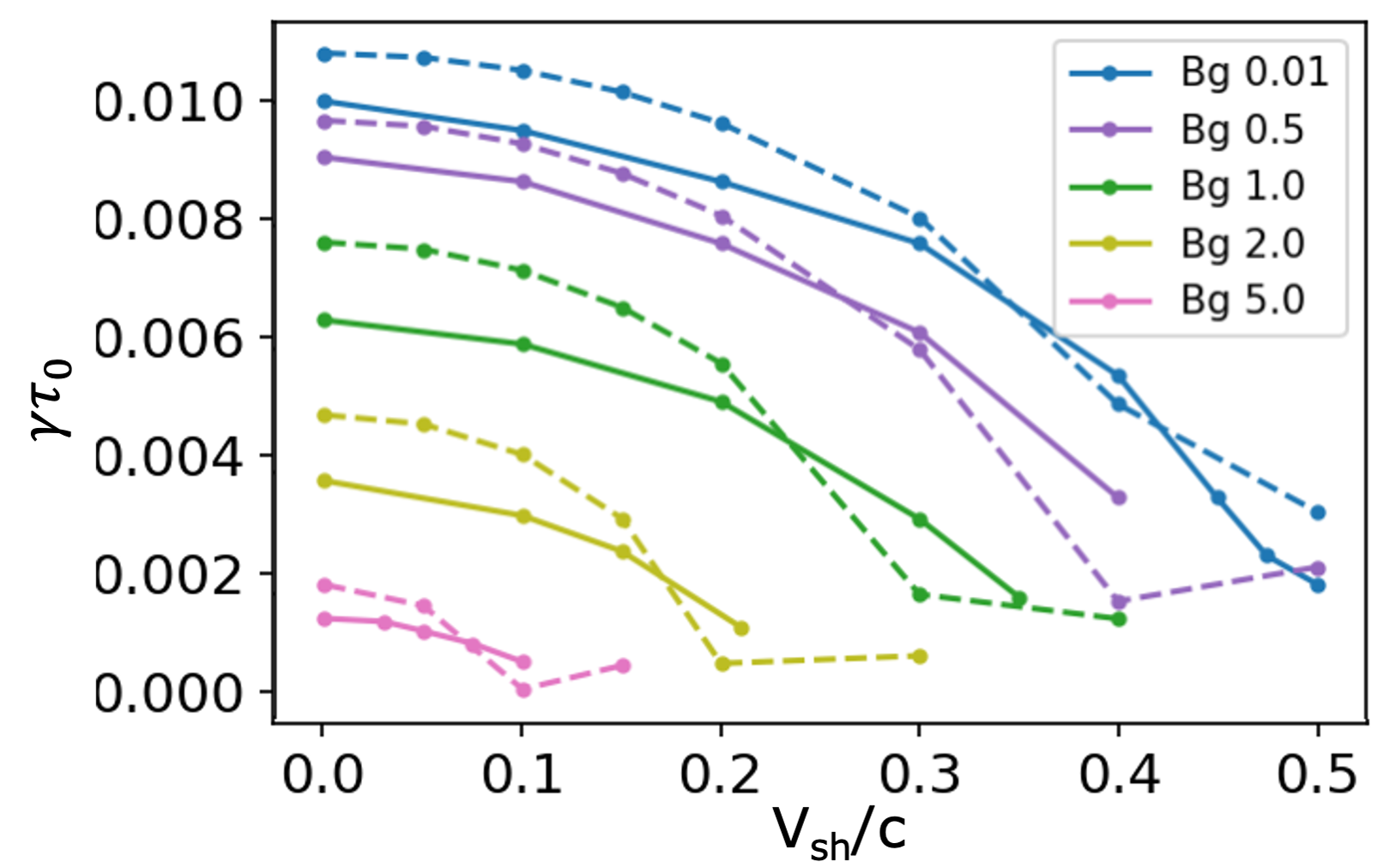}
  \caption{The average growth rate for the perturbation runs (solid lines) with $B_g/B_{x0}= 0.01, 0.5, 1,2,5$ and $V_{sh}$ varied from $0 \sim 0.5c$ plotted with the solver growth rate (dashed lines).} 
  \label{eight} 
\end{figure}

We would like to stress that the linear tearing analysis can still be useful for simulations initiated with a magnetic perturbation, such as those of \cite{peery}, which may not have a clear linear phase but still exhibit delay in the onset of reconnection due to the shear flow.
In these simulations, a thick current sheet is used so onset will necessarily be preceded by thinning of the current sheet, which is known to happen concurrently with linear tearing \cite{honshino21}. Average growth rates can be found from these simulations as well, as the inverse of the reconnection onset time. To compare the simulations to the solver results, we use an effective current sheet width, which takes into account the enhancement from the initial perturbation. Using this, the solver predicted growth rates can be shown to match well with the scaling of the simulation averages. Fig. \ref{eight} shows the average growth rates from simulations with a thick current sheet and perturbation for five values of guide field, as a function of shear flow, compared to solver results using $L_B=2.5d_e$.
The scaling we see here indicates that the shear flow also affects the buildup phase of these simulations, and can be modeled using standard tearing theory, even if we cannot see direct evidence of an initial linear phase.

\subsection {Summary}
We have found that increasing shear flow speed, in the direction of the reconnection field, delays the onset of reconnection. To explain this, we present numerical model for the linear phase of the tearing instability in the relativistic regime. We use fluid equations for a pair plasma \cite{Koide9} and include for the first time, the effects of the motional electric field, which is not negligible when the guide field is finite. Consistent with previous studies of tearing instability, we find that the growth in the linear phase is slowed by shear flow, leading to a long buildup phase before magnetic reconnection can onset.
The solver also allows for the exploration of the transition from tearing dominated growth at low shear flow speeds to the Kelvin Helmholtz instability dominated regime as shear flow speeds are increased. We find that there is an intermediate regime where characteristics of both KHI and tearing instability can be observed and reconnection can still onset due to current sheet deformation. 

The numerical solver was validated by comparison to previous literature, detailed in Appendix C, and to growth rates calculated from VPIC simulations, which show the same scaling. 
The growth rate of linear tearing decreases with increasing shear flow speed, wider initial current sheet width, or stronger guide field strength (consistent with the decrease in the in-plane Alfv\'en velocity). Increasing the reconnecting field strength increases the growth rate initially, but the effects saturate at high enough field strength.

Understanding where reconnection will onset is important to many systems in both non-relativistic and relativistic regimes. Of highest relevance to this study, reconnection can be responsible for flux depletion and high energy particle acceleration from astrophysical sources \cite{sironi_21, kirk_03, lyutikov_03,  benoit_12}. Therefore, a theoretical understanding of the dynamics of reconnection can be valuable to understanding observations which cannot be replicated to test on Earth.

Additional future work could include extending the study to three dimensions. Theoretical studies allowing for oblique modes find that the growth rate can be modified \cite{baalrud, miura_82} and 3D simulations can include instabilities which do not appear when there is no variation in the out-of-plane direction \cite{zenitani_09}. Also, further characterization of the relativistic KHI and intermediate regime which is seen in the solver and simulation but not discussed in depth here, could be conducted.

\onecolumngrid
\appendix*
\section{A (Main Solver)}
In this section we show details on the numerical solver, starting from the relativistic equations for a pair plasma \cite{Koide9},
\beq
\label{a1}
 h \partial_\nu (U^\nu \vb{U})+h_J \partial_\nu (J^\nu \vb{J}) 
= (\nabla \times \vb{B}- \partial_t \vb{E}) \times \vb{B} + \rho_q \Gamma \vb{E}
\eeq
\beq
\label{a2}
\Gamma \vb{E}= -\vb{U} \times \vb{B}+ S^{-1} (\vb{J}+ \Gamma \rho_q \vb{U})]
+ h_J \partial_\nu (J^\nu \vb{U}+ U^\nu \vb{J}).
\eeq
where $h_J=h/4n^2$ and $h, n$ are the normalized simulation quantities $n=1$. Enthalpy is defined $h=(n m_e c^2+\frac{5}{2}P)$ so for a temperature $T=0.5 mc^2$, as in the simulation, we have a normalized value of $h=9/4 n$.

Given the forms of Eqs. (10)-(15), the linear forms of derived quantities $\vb{E}, \vb{J},$ and $ \rho_q$ in terms of $\tilde{\psi}$ and $\tilde{\phi}$ are: \\

\noindent $\tilde{E}_x=(\tilde{U}_z/ \Gamma_0) B_{g}$\\
$\tilde{E_z}= - (1/\Gamma_0) B_g \tilde{U}_x+U_{x0}B_g(\tilde{\Gamma}/\Gamma_0^2)$\\
$\tilde{E}_y= \gamma \tilde{\psi}$\\
$\tilde{J_x}= -\partial_t \tilde{E_x}$\\
$\tilde{J}_y=-\partial_x \tilde{B}_z + \partial_z \tilde{B}_x $\\
$\tilde{J}_z= -\partial_t \tilde{E}_z$\\
$\tilde{\rho_q}= \partial_x \tilde{E}_x + \partial_z \tilde{E}_z$\\

\noindent where $\tilde{E}_x, \tilde{E}_z$ are found from ideal Ohms law, $\tilde{E}_y$ is found from Faraday's law, $\tilde{\vb{J}}$ is found from Amp\'eres law and $\tilde{\rho}_q$ is found from Gauss's law. 
\noindent When these forms, as well as the forms of Eqs. (10)-(15) are substituted into the momentum equation and Ohms law, keeping only the linear term in the expansion, the final equations can be written as the generalized eigensystem;
\beq
\label{eig}
\gamma \begin{pmatrix}
A_1 & A_2 \\
A_3 & A_4 \\
\end{pmatrix}
\begin{pmatrix}
\tilde{\psi}\\
\tilde{\phi} \\
\end{pmatrix}=
\begin{pmatrix}
B_1 & B_2\\
B_3 & B_4 \\
\end{pmatrix}
\begin{pmatrix}
\tilde{\psi}\\
\tilde{\phi}\\
\end{pmatrix}
\eeq
where the coefficients $A_1, A_2, B_1, B_2$ come from Ohms law, and $A_3,A_4, B_3, B_4$ come from the momentum equation.  In the following we use $\reflectbox{L}_0=(1/\Gamma_0)$ as an abbreviation. The coefficients are given;\\

\noindent$A_1=\Gamma_0- h_J \Gamma_0(\partial_z^2-k^2)$\\
$A_2=-h_J J_{y0} U_{x0} \reflectbox{L}_0\partial_z$\\
$B_1=-i k U_{x0}+h_J i k U_{x0}(\partial_z^2-k^2) +S^{-1} (\partial_z^2-k^2)$\\
$B_2=i k B_{x0}-h_J i k J_{y0}'$\\

\noindent$A_3=0$\\
$A_4= m1+m3+m4+m6$\\
$B_3= m5$\\
$B_4=m2+m5+m7$\\

\noindent The coefficients $m1-m7$ are derived as the following:\\

$m1= \gamma h [\Gamma_0 (-k^2 \tilde{\phi} +\tilde{\phi}''+\Gamma_0' \tilde{\phi}')+ U_{x0}^2 \reflectbox{L}_0 \tilde{\phi}''+(2 U_{x0} U_{x0}' \reflectbox{L}_0+ U_{x0}^2 \reflectbox{L}_0' )\tilde{\phi}']$\\

$m2= h[ i k U_{x0}(\tilde{\phi}''-k^2\tilde{\phi})- i k \tilde{\phi} U_{x0}'']$\\

$m3= \gamma h_J \{(E_{z0}' J_{x0} U_{x0} \reflectbox{L}_0)' \tilde{\phi}'+E_{z0}' J_{x0} U_{x0} \reflectbox{L}_0 \tilde{\phi}''
+ (\Gamma_0 J_{x0})'[ B_g \reflectbox{L}_0(k^2 \tilde{\phi}- \tilde{\phi}'')- (B_g \reflectbox{L}_0)' \tilde{\phi}'- E_{z0}U_{x0}\reflectbox{L}_0^2 \tilde{\phi}''-(E_{z0}U_{x0}\reflectbox{L}_0^2)' \tilde{\phi}']
+\Gamma_0 J_{x0}[k^2(B_g \reflectbox{L}_0)' \tilde{\phi}+k^2B_g \reflectbox{L}_0 \tilde{\phi}'-2(B_g \reflectbox{L}_0)' \tilde{\phi}''-B_g \reflectbox{L}_0 \tilde{\phi}''' 
-(B_g \reflectbox{L}_0)'' \tilde{\phi}'-2(E_{z0}U_{x0}\reflectbox{L}_0^2)' \tilde{\phi}''-E_{z0}U_{x0}\reflectbox{L}_0^2 \tilde{\phi}'''-(E_{z0}U_{x0}\reflectbox{L}_0^2)'' \tilde{\phi}']\}$\\

$m4= h_J \gamma \{(k^2J_{x0}+J_{x0}'')(B_g\reflectbox{L}_0 \tilde{\phi}' +E_{z0}\reflectbox{L}_0^2 U_{x0} \tilde{\phi}')- 2k^2 (J_{x0} B_g \reflectbox{L}_0)' \tilde{\phi}-2 k^2 J_{x0} B_g \reflectbox{L}_0 \tilde{\phi}'
 +2 J_{x0}'[B_g\reflectbox{L}_0 \tilde{\phi}'' +(B_g\reflectbox{L}_0)' \tilde{\phi}'+E_{z0}\reflectbox{L}_0^2 U_{x0} \tilde{\phi}''+(E_{z0}\reflectbox{L}_0^2 U_{x0})' \tilde{\phi}']
 +J_{x0}[
2(B_g\reflectbox{L}_0)' \tilde{\phi}'' +(B_g\reflectbox{L}_0)'' \tilde{\phi}'+B_g\reflectbox{L}_0 \tilde{\phi}'''+2(E_{z0}\reflectbox{L}_0^2 U_{x0})' \tilde{\phi}''+(E_{z0}\reflectbox{L}_0^2 U_{x0})'' \tilde{\phi}'+E_{z0}\reflectbox{L}_0^2 U_{x0} \tilde{\phi}''']\}$\\

$m5= h[ i k B_{x0}(\tilde{\psi}''-k^2\tilde{\psi})- i k \tilde{\psi} B_{x0}'']$\\

$m6=\gamma B_g[B_g \reflectbox{L}_0(k^2 \tilde{\phi}- \tilde{\phi}'')- (B_g \reflectbox{L}_0)' \tilde{\phi}'- E_{z0}U_{x0}\reflectbox{L}_0^2 \tilde{\phi}''-(E_{z0}U_{x0}\reflectbox{L}_0^2)' \tilde{\phi}']- B_g'[B_g\reflectbox{L}_0 \tilde{\phi}' +E_{z0}\reflectbox{L}_0^2 U_{x0} \tilde{\phi}] $\\

$m7=-ik \Gamma_0 E_{z0}[B_g \reflectbox{L}_0(k^2 \tilde{\phi}- \tilde{\phi}'')- (B_g \reflectbox{L}_0)' \tilde{\phi}'- E_{z0}U_{x0}\reflectbox{L}_0^2 \tilde{\phi}''-(E_{z0}U_{x0}\reflectbox{L}_0^2)' \tilde{\phi}']
 -i k[E_{z0}''B_g \reflectbox{L}_0 \tilde{\phi}+E_{z0}'(B_g \reflectbox{L}_0)' \tilde{\phi}]$\\

\noindent These coefficients will each be N $\times$ N matrices where the diagonal elements are given by the above functions of $z$ and N is the resolution. Any off-diagonal terms come from the derivatives, which are written as centered second order finite difference approximations,
\beq
\partial_z=(8*\vb{I}_{i=1}-8*\vb{I}_{i=-1}-\vb{I}_{i=2}+\vb{I}_{i=2})/12d
\eeq
\beq
\partial_z^2=(-\vb{I}-30+16\vb{I}_{i=1}+16\vb{I}_{i=-1}-\vb{I}_{i=2}-\vb{I}_{i=2})/12 d^2
\eeq
where $d$ is the step size and $\vb{I}_{i=1}$ indicates an identity matrix with an diagonal offset by $i$.

\noindent The non-relativistic limit of these equations recover the linear equations of Faganallo \cite{faganello}:\\

\noindent$A_1 = 1-d_e^2(\partial_z-k^2)$\\
$A_2, A_3=0$\\
$A_4=(\partial_z-k^2)$\\

\noindent $B_1 =(S^{-1}+\tfrac{1}{2} d_e ikV_{x0})(\partial_z-k^2)- ikV_{x0}$ \\
$B_2 = ikB_{x0}-\tfrac{1}{2}d_e^2 ik B_{x0}''$\\
$ B_3 =i k B_{x0}(\partial_z-k^2)- i k B_{x0}''$\\
$ B_4 = -i k V_{x0}(\partial_z-k^2) +  i k V_{x0}''$\\

\noindent In these equations the electron inertial length $d_e$ parametrizes the strength of the inertial non-ideal terms and the non-relativistic Lundquist number is used $S^{-1}=L_B V_a/ \eta$. The normalization is done in the standard non-relativistic convention as in \cite{faganello, chenmorrison, MacTaggart2020, betar20}, with respect to the non-relativistic $V_a= B/\sqrt{4 \pi \rho}$, the current sheet width $L_B$, and the total magnetic field magnitude $|B|$. So growth rates are presented as $\gamma \tau_A$. The guide field is assumed to be uniform and its strength enters the equation only through the normalization of the reconnecting magnetic field.

\section{B (Jet)}
In this section we give details on the derivation of $Jet$ the numerical solver which uses two-fluid contributions to the Lorentz factor, in the limit of no initial shear flow ($V_{sh}=U_{x0}=0$).
The initial magnetic fields and currents remain the same as above, as do the forms of the perturbed velocity and magnetic fields, however the motional electric field will now be zero; ($\vb{U}_0=\vb{E}_0=\rho_{q0}=0$).
And the Lorentz factor will be found using the definition;
\begin{equation}
\Gamma=U^0=\frac{1}{n} \left(n_e \sqrt{1+U^2_e} + n_i \sqrt{1+U^2_i} \right)
\end{equation}
where $n$ is the total density. The single fluid initial velocity is zero but the two-fluid velocities will have contribution from $J_{x}$ and $J_y$. We assume these to be evenly split between positions and electrons.
Therefore, to find $\tilde{\Gamma}$ we must consider the contributions from the perturbed current;
\beq
\begin{split}
\tilde{\Gamma}= &\frac{1}{n} \left[ \frac{n_e}{\Gamma_{0e}}(U_{x0e}\tilde{U}_{xe}+U_{y0e}\tilde{U}_{y0e})+\frac{n_i}{\Gamma_{0i}}(U_{x0i}\tilde{U}_{xi}+U_{y0i}\tilde{U}_{y0i}) \right]\\
&= G \frac{1}{4} \left[i k \gamma  J_{x0} B_g \reflectbox{L}_0 \tilde{\phi}+ J_{y0} (\tilde{\psi}''-k^2 \tilde{\psi}) \right]+ g J_{x0} \tilde{\phi}'
\end{split}
\eeq

where $G=\frac{1}{2} \left( \frac{1}{\Gamma_{0e}}+ \frac{1}{\Gamma_{0i}} \right)$ and $g=\frac{1}{4} \left( \frac{1}{\Gamma_{0i}}- \frac{1}{\Gamma_{0e}} \right)$.

Without $U_{x0}$, we have $\tilde{E}_z=-(1/\Gamma_0) B_g \tilde{U}_x$ which simplifies $\tilde{J}_z$ and the charge separation $\tilde{\rho}_q$. The rest of the derived perturbed quantities remain the same. Plugging these definitions into eqs (\ref{a1}), (\ref{a2}) and linearizing in the standard way we find the coefficients:\\

\noindent$A_1=\Gamma_0- h_J \Gamma_0(\partial_z^2-k^2)+ G J_{y0}^2 \frac{1}{4}(\partial_z^2-k^2)$\\
$A_2= \gamma h_JJ_{y0} g J_{x0}$\\
$B_1=S^{-1} (\partial_z^2-k^2)$\\
$B_2=i k B_{x0}-h_J i k J_{y0}'$\\

\noindent$A_3=0$\\
$A_4= m1+m3+m4+m5$\\
$B_3= m2$\\
$B_4=0$\\

\noindent where $m1-m5$ are defined:\\

$m1= \gamma h [\Gamma_0 (-k^2 \tilde{\phi} +\tilde{\phi}''+\Gamma_0' \tilde{\phi}')$\\

$m2= h[ i k B_{x0}(\tilde{\psi}''-k^2\tilde{\psi})- i k \tilde{\psi} B_{x0}'']$\\

$m3=\gamma B_g[B_g \reflectbox{L}_0(-k^2 \tilde{\phi}+ \tilde{\phi}'')+ (B_g \reflectbox{L}_0)' \tilde{\phi}'+ B_g'\reflectbox{L}_0 \tilde{\phi}']$\\

$m4=\gamma h_J[ -\Gamma_0 J_{x0} (B_g\reflectbox{L}_0 \tilde{\phi}'''+ 2( B_g \reflectbox{L}_0)' \tilde{\phi}'+ (B_g \reflectbox{L}_0)'' \tilde{\phi})- (\Gamma_0 J_{x0})'(B_g \reflectbox{L}_0 \tilde{\phi}''+ (B_g \reflectbox{L}_0)' \tilde{\phi}')]$\\

$m5= h_J \gamma \{(k^2J_{x0}+J_{x0}'')(B_g\reflectbox{L}_0 \tilde{\phi}')- 2k^2 (J_{x0} B_g \reflectbox{L}_0)' \tilde{\phi}-2 k^2 J_{x0} B_g \reflectbox{L}_0 \tilde{\phi}'
 +2 J_{x0}'[B_g\reflectbox{L}_0 \tilde{\phi}'' +(B_g\reflectbox{L}_0)' \tilde{\phi}']
 +J_{x0}[
2(B_g\reflectbox{L}_0)' \tilde{\phi}'' +(B_g\reflectbox{L}_0)'' \tilde{\phi}'+B_g\reflectbox{L}_0 \tilde{\phi}''']\}$\\

This system is solved in the same manner as the main solver defined in Appendix A.

\section{C (Solver Benchmarking)}
\subsection{Tearing Instability}

\begin{figure*}[ht]
  \centering
  \includegraphics[width=\linewidth]{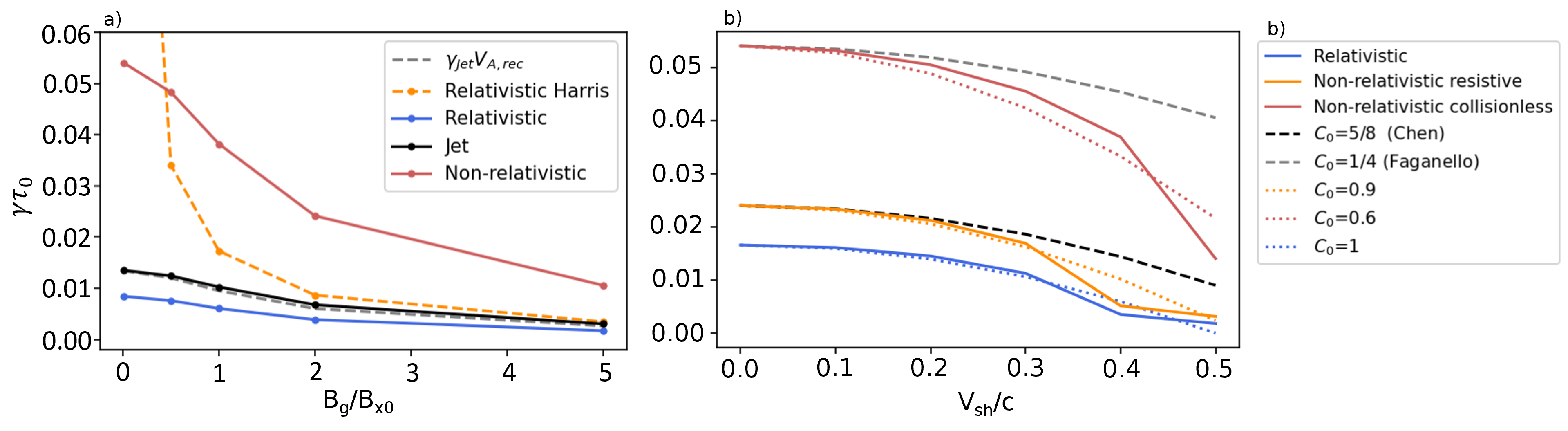}
  \caption{Panel a) scaling of the inertial tearing growth rate with guide field for; the relativistic solver using a simple Harris sheet (orange), the main relativistic solver (normalized to $t_0=\tau_{pe}$) with a force free current sheet (blue), $Jet$ the solver with two-fluid $\Gamma$ (black), and the non-relativistic solver (normalized to $t_0=\tau_A$) (red). And in dashed gray scaling of $(\gamma_0 \tau_0) V_{A, rec}$ compared to Jet; Panel b) shows scaling of the growth rate with $V_{sh}$ for the non-relativistic resistive, non-relativistic inertial, and full relativistic (inertial) tearing. Dashed lines show Eq. (\ref{cm5}) with $C_0$ given by \cite{chenmorrison, faganello}. Dotted lines show best fit $C_0$ for the solver.}
  \label{lines} 
\end{figure*}

As validation of the solver in this section we compare to some familiar limits in literature. 
Figure \ref{lines} shows scaling results for $B_g=B_{x0}$, in both the relativistic and the non-relativistic regimes, which may be compared to the analytic growth rates derived in multiple previous studies of linear tearing. For that reason non-relativistic results shown in this plot use a simple Harris sheet, while relativistic results use the more accurate force free initial conditions.

In Figure \ref{lines}a, we return to the case of $V_{sh}=0$, as in Fig. \ref{jet}.
Besides $Jet$ and the main relativistic solver we show results of the relativistic solver with a simple Harris sheet; $\vb{B}=\tanh(z/L_B) \hat{x} +B_g \hat{y}$, where the guide field is uniform. We also show results in the non-relativistic limit using the equations of \cite{faganello} as discussed in Appendix A.
These are also calculated with a simple Harris sheet, to compare to previous studies. We present non-relativistic and relativistic growths on the same plots for comparison but note that the relativistic growth is normalized to $\tau_A$ and not $\tau_{pe}$ as the relativistic solver and simulations are, as we are interested in comparing scaling behavior only \cite{faganello, komissarov, chenmorrison, MacTaggart_2018, barkov}.

In the non-relativistic linearized equations, the magnitude of the growth varies with the strength of the reconnecting magnetic field, therefore the effect of the guide field is felt through the normalization of the initial reconnecting field, $B_{x0, norm}=1/\sqrt{1-(B_g/B_{x0})^2}$.
In the relativistic regime (with force free initial conditions) we find that the effect of a guide field is weaker, which is to be expected since $B_y$ will be non-zero at the current sheet in all cases. With the simple Harris current sheet, the case with no guide field finds a drastically faster growth rate. This is because, especially at high magnetization, the simple sheet doesn't provide a valid force balance since it does not include pressure enhancement at the current sheet. 
This demonstrates the necessity of using the relativistic force free initial conditions. 
A decrease of the growth rate with guide field agrees with work on relativistic tearing \cite{zenitani_08, barkov, Cerutti_2014, Demidov_2025, barkov, yang17}. 
For resistive tearing modification of the growth rate is found to be $\gamma_{B_g} \sim \gamma_0 \sqrt{V_{A, rec}}$ \cite{yang17}, though we find the scaling (for the inertial case shown here) to be close to $ \sim \gamma_0 V_{A, rec}$.



In Fig. \ref{lines}b we allow $V_{sh}$ to vary, keeping $B_g=0.01$ and $kd_e=0.2$ constant and we plot the growth rate as a function of shear flow speed. 
We find that when $V_{sh} \sim 0.5 V_A$  the Kelvin-Helmholtz instability begins to dominate, so speeds above that are not shown here.
In all three cases the growth rate decreases with the flow speed with a similar scaling to non-relativistic analytical relations given by \cite{faganello, chenmorrison, Hofman_1975};
\beq
\label{cm5}
\gamma \sim \gamma_{(V_{sh}=0)}\bigg[1- C_0 \left(\frac{G(0)'} {F(0)'}\right)^2 \bigg]
\eeq
where $F(z)=\vb{B}_{0}(z)\cdot \hat{k}/|B|$ and $G(z)=\vb{V}_{sh}(z) \cdot \hat{k}/V_A$. The constant $C_0$ is given in each paper as it  depends on the initial setup and assumptions made. $C_0= 1/4$ for inertial tearing and $C_0=5/8$ for resistive tearing. We find empirically that the growth calculated by the solver 
has different coefficients of best fit. The doted lines are plotted for $C_0=0.6$ in the inertial case and $C_0=0.9$ in the resistive case. We expect this discrepancy between the solver and the analytical scaling due to differing approximations in the analytical treatment of \cite{chenmorrison, faganello}. 
The scaling of the relativistic solver is also shown, plotted with an analytical scaling using $C_0=1$. Even though this is a non-relativistic equation, it fits well especially for low values of shear flow.

\subsection{Kelvin Helmholtz Instability}
While we focus mainly on the tearing instability in this work, it is useful to discuss the transition between to KHI, especially as many previous studies of relativistic KHI focus on uniform magnetization \cite{chow23, ferrari, osmanov}.
\begin{figure}[h!]
  \centering
  \includegraphics[width=\linewidth]{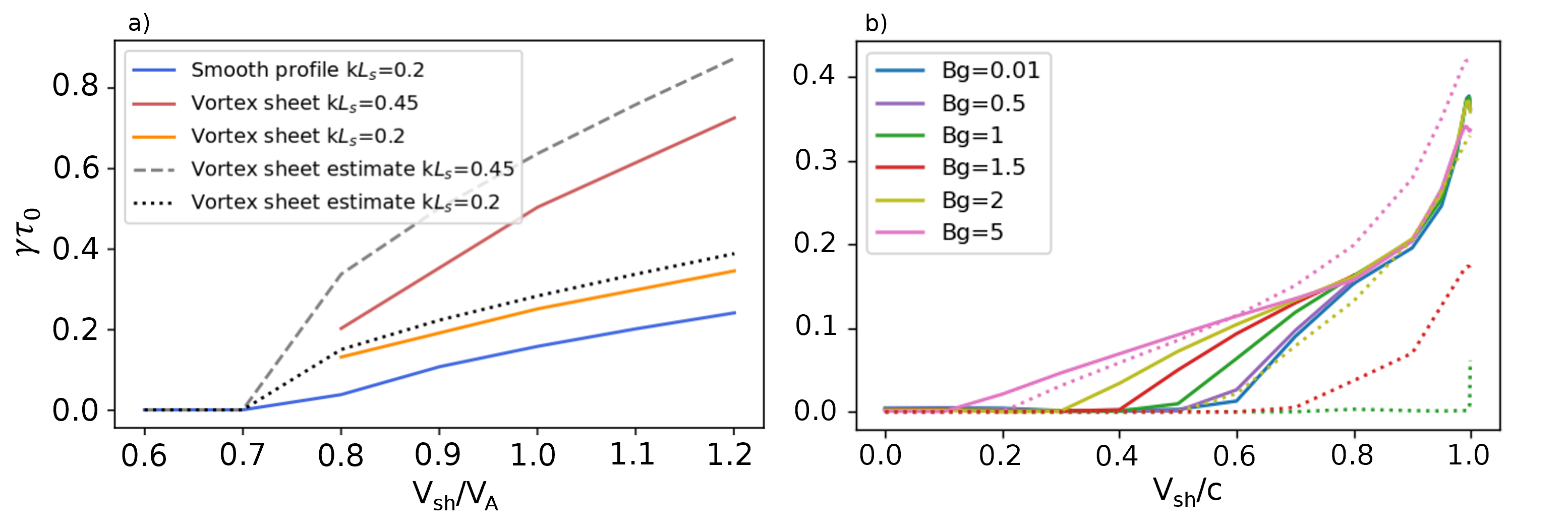}
  \caption{Panel a) shows scaling of KHI with uniform $B_{x0}$ for a smooth velocity profile with $k L_{s}=0.2$, (blue line),  vortex sheets with $kL_{s}=0.45$ (red line) and $kL_{s}=0.2$ (orange line), and analytical estimates for a vortex sheet Eq. (\ref{four4}) \cite{CHANDRA}; panel b) shows relativistic growth at a force free current sheet (solid lines) and in a uniform field (dotted lines).} 
  \label{khis} 
\end{figure}

Using the same non-relativistic equations as above (those of \cite{faganello}), we can recover the non-relativistic, incompressible limit. 
Fig. \ref{khis}a shows the growth rates from the solver with uniform magnetization to compare to the well-known dispersion relation for a discontinuous velocity profile (called a vortex sheet),simplified here as we assume uniform density and antisymmetric shear in the magnetic field and velocity \cite{CHANDRA}; 
\beq
\label{four4}
(\gamma \tau_A)^2=[\vb{k} \cdot \vb{V}_{sh}]^2/V_A- [\vb{k} \cdot \vb{V}_A]^2/V_A
\eeq
Here $\tau_A=L_{s}/V_A$ where $L_s$ is the shear width. 
We plot results for $B_{x0}=B_g$ at two different wavenumbers $k L_{s}=0.45$ and $kL_{s}=0.2$. To approximate a vortex sheet in the solver we use a small $L/L_{s}=0.06$ and we can see that while there is reasonable comparison at the smaller wavenumber, Eq. \ref{four4} breaks down at higher wavenumbers \cite{miura_82, ONG19721}.
We also plot the solver solution for a smooth shear profile with $kL_{s}=0.2$, for comparison to the vortex sheet approximation. 
In all cases the growth rate scales linearly with increased shear speed, as expected.


Now we consider the effects of a current sheet in the relativistic regime. We use the full equations for our main solver. Fig. \ref{khis}b  shows the growth rate for relativistic KHI for five values of guide field. Note that the shear flow is now normalized to the speed of light and not the Alfv\'en velocity. Solid lines show the growth rate found with the force free current sheet, and dashed lines show cases with uniform magnetic fields. All were found for $kL_s=0.15$.
At a current sheet, KHI becomes unstable at a lower shear speed and the growth rate is enhanced, due to the in-plane magnetic depletion \cite{KEPPENS}.
Tearing is also unstable here, but hard to identity visually as $\gamma_{tear} \ll \gamma_{KHI}$.

In the relativistic regime the growth rate of KHI still scales with the shear flow speed, however as $V_{sh}$ approaches the speed of light, it loses linearity.
This may be a nonphysical behavior; previous studies expect that at super-relativistic speeds the growth rate is suppressed \cite{hamlin_13, chow23, osmanov} because
the inertia of the enthalpy perturbations becomes too large. Since we have assumed incompressibility and uniform enthalpy we cannot capture this effect, and thus the assumptions used in our linearized equations are not valid close to the speed of light.
Even so we may use the solver to examine some limiting behavior.
It has been a question of interest whether the KHI will be completely stabilized in highly magnetized plasma where $V_{A} \rightarrow c$,  and we see that this may be the case, if the magnetic field is predominantly parallel to the wave vector. 
However in the presence of a current sheet and a guide field, KH modes are unstable at much lower velocities, even where the Alfv\'en velocity is large. 


\section{D (VPIC setup)}

In this work, we validate our model against 2D Particle-in-Cell (PIC) simulations using VPIC code \citep{vpic}, which solves the fully relativistic dynamics of plasma and electromagnetic fields in flat space. The initial conditions are a force-free current sheet \citep{guo_14, ys19} with uniform background density and temperature in the proper frame.  We employ electron-positron pair plasma, with the particle mass of $m_i=m_e \equiv m$ and temperature ratio $T_p/T_e=1$, motivated by studies that argue pair plasma are relevant in astrophysical plasmas \citep{Arons_12,barniol_17}. 

The initial reconnecting field is ${\bf B}= B_{x0} \tanh(z/L) \hat{x}+ B_y(z) \hat{y}$. An in-plane shear flow is implemented as ${\bf V}=V_{sh} \tanh(z/L_s) \hat{x}$, which is accompanied by the motional electric field $E_z=-V_x B_y/c$. We require that the guide field have asymptotic value $B_g$, so the spatial dependence of $B_y(z)$ is found from the force free condition on the pressure balance $B_x^2+ B_y^2-E_z^2=B_{x0}^2+ B_g^2-(V_{sh} B_g/c)^2$. 
The current density $J$ is found from Amp\`ere's law, to satisfy the spatial variation in $B_x$ and $B_z$. It is implemented assuming that electrons and positrons each symmetrically carry half the current and half the charge.
The density profile in the simulation frame is $n= \gamma_d n_{0}+ \delta n$ where the charge separation is added assuming $\delta n_i= -\delta n_e= \rho_c/2$ with charge density $\rho_c$ found from Gauss' law. The Lorentz factor is found from the initial drift of the current, $\gamma_d=\sqrt{1+J^2/(4n_0^2e^2c^2)}$.

Multiple initial current sheet widths were used. The initial setup of \cite{peery} used a thick current sheet of $L_B= 10 d_e$, and to induce reconnection X-line at the domain center, a magnetic perturbation of strength $\delta B_z=0.13 B_{x0}$ was used in the initial condition. For study of the linear tearing phase the perturbation is not used, instead thinner current sheets of $L_B= 2, 4 d_e$ are implemented. The velocity shear width was kept at $L_s=0.5L_B$ in all cases.
Two sets of domain sizes were used. Lower resolution simulations were run with system size is $L_x \times L_z= 384 d_e\times 384 d_e$, with $ 2048 \times 2048$ grid points and 100 macroparticles per cell. Higher resolution simulations were run with box size $L_x \times L_z= 150 d_e\times 130 d_e$ and resolutions as listed in fig. \ref{table}.
Tests were run with larger box sizes and it was found that for the linear and starting growth phases of reconnection, the small box does not affect evolution. Later stages of reconnection may see effects, but were not considered in this paper.
The boundary conditions are periodic in the x-direction, reflecting for particles and conducting for the fields in the z-direction.

In this paper, density is normalized to $n_0=1$, velocities are normalized to the speed of light $c$ and length scales are normalized to the inertial length $d_e= c/\omega_{pe}$, such that time scales are normalized to the plasma frequency $\omega_{pe}=\sqrt{ 4 \pi n_0 e^2/ m}= 1$, (note $\omega_{pe}=\omega_{pi}$).
The temperature in the bulk flow frame for both species is $T_0=0.5mc^2$.
For older runs $\omega_{pe}/\Omega_{ce}=0.0666$, where $\Omega_{ce}= e B_0/mc$ is the ion cyclotron frequency. The asymptotic magnetization is then $\sigma_{x0}=B_{x0}^2/[n_e m_e c^2+ \gamma/(\gamma-1)P_e]= 100$ for the specific heat ratio $\gamma=5/3$ and enthalpy $h_0\simeq 2.25$. 

\bibliography{ref}{}
\bibliographystyle{aasjournal}
\end{document}